\documentclass[a4paper,fleqn]{cas-dc}

\usepackage[authoryear]{natbib}
\setcitestyle{maxnames=2,minnames=1}

\usepackage{booktabs}
\usepackage{tabularx}
\usepackage{array}
\usepackage{multirow}
\usepackage{makecell}
\usepackage{makecell}
\usepackage{xcolor} 

\newcolumntype{Y}{>{\raggedright\arraybackslash}X}

\begin{document}

\title{What Challenges Do Developers Face in AI Agent Systems?
An Empirical Study on Stack Overflow \& GitHub Issues}
\shorttitle{Developer Challenges in AI Agent Systems}
\shortauthors{Asgari et al.}
\author[1]{Ali Asgari}
\ead{a.asgari-2@tudelft.nl}

\author[1]{Annibale Panichella}
\ead{a.panichella@tudelft.nl}

\author[2]{Pouria Derakhshanfar}
\ead{pouria.derakhshanfar@jetbrains.com}

\author[1]{Mitchell Olsthoorn}
\ead{m.j.g.olsthoorn@tudelft.nl}

\affiliation[1]{organization={Delft University of Technology},
  city={Delft}, country={The Netherlands}}

\affiliation[2]{organization={JetBrains Research},
  city={Amsterdam}, country={The Netherlands}}

\begin{abstract}
\emph{Context:} AI Agents have rapidly gained prominence in both research and industry as systems that extend large language models with planning, tool use, memory, and goal-directed action. Despite this progress, the development and maintenance of Agent systems present recurring engineering difficulties that are not yet well characterized in developer-facing evidence.\\ 
\emph{Objective:} To address this gap, this study analyzes developer discussions on Stack Overflow and failure reports from GitHub issue trackers associated with widely used Agent frameworks. \\
\emph{Method:} For Stack Overflow, an Agent-focused corpus is constructed through tag expansion and filtering, latent themes are derived using LDA-MALLET, and topics are manually validated and labeled. For GitHub, a taxonomy of issue themes is developed to capture deployment-time failures and maintenance burdens. \\
\emph{Results:} Analysis across both platforms identifies seven Stack Overflow topics (comprising 28 subtopics) and thirteen GitHub issue topics, which are synthesized into five overarching families of \textbf{major Agent challenges: (1) environment, platforms, and dependency management; (2) retrieval, embeddings, and Agent memory; (3) orchestration and execution control; (4) interaction contracts between models and tools; and (5) runtime reliability and operational robustness}. Topic popularity and difficulty are quantified, revealing that widely discussed issues, such as installation and prompting, are often resolved more quickly, whereas retrieval- and orchestration-related challenges are less visible, more complex, and tend to persist as ongoing maintenance burdens on GitHub.
\end{abstract}


\maketitle
\section{Introduction}

AI Agents have recently emerged as a prominent paradigm for constructing AI-enabled systems that extend large language models with capabilities such as tool use, memory, planning, and multi-step execution. These systems are increasingly adopted in both research prototypes and production applications, supported by rapidly evolving frameworks and platforms. However, developers continue to encounter significant practical challenges in building, deploying, and maintaining Agent-based systems.

Although prior work has examined AI Agents through surveys, conceptual frameworks, and system descriptions, there is limited empirical evidence regarding the practical challenges developers encounter at scale. Much of the existing literature is based on author-driven experiences or controlled evaluations, rather than large-scale, developer-reported data \cite{cemri2025multi,deng2025ai,zou2025security,de2025open,gabriel2025we,kolt2025governing,tian2025outlook,du2021survey,du2025ai,sapkota2025ai}. Consequently, the most prevalent and difficult challenges, as well as their distribution across the Agent ecosystem, remain unclear.

To address this gap, this study analyzes discussions from Stack Overflow, a large developer-oriented Q\&A platform with approximately 60 million questions and answers and around 30 million registered users. Stack Overflow has been widely utilized in empirical software engineering research across domains such as Explainable AI \cite{sayyadnejad2024exploring}, concurrency \cite{ahmed2018concurrency}, GPU programming \cite{yang2023understanding}, Ruby programming~\cite{akbarpour2025unveiling}, and deep learning~\cite{han2020programmers}. Here, Stack Overflow is leveraged as a high-signal source of real-world developer challenges related to AI Agent development.

The objective of this study is to systematically characterize the technical and practical challenges developers encounter when working with AI Agents. The research is structured around the following research questions:
\begin{enumerate}
    \item[\textbf{RQ1}:] \textit{What topics do AI Agent developers ask about?}
    \item[\textbf{RQ2}:] \textit{Which topics are the most popular and the most difficult?}
    \item[\textbf{RQ3}:] \textit{How do challenge signatures differ across AI Agent ecosystem domains?}
\end{enumerate}

To address \textbf{RQ1}, Stack Overflow was mined at scale to construct an Agent-focused corpus through tag expansion and filtering, resulting in 3,191 unique questions with accepted answers. Topic modeling was performed using LDA-MALLET~\cite{mccallum2002mallet}, producing a taxonomy of seven high-level topics and 28 subtopics. A statistically representative sample of 343 questions was selected using Cochran’s sample size formula with finite-population correction~\cite{woolson1986sample} and Neyman allocation~\cite{neyman1938contribution} for manual validation. In parallel, 64,098 GitHub issues were extracted from 18 widely used AI Agent repositories. The same LDA-MALLET pipeline was applied to derive a complementary taxonomy of 13 issue topics, capturing implementation, integration, and maintenance failures in deployed Agent systems. These topics were further validated through manual analysis of a stratified sample of 382 GitHub issues, selected using Neyman allocation to preserve proportional topic representation. Finally, insights from Stack Overflow and GitHub were synthesized into five higher-level families of major Agent-specific challenges: (1) environment, platforms, and dependency management; (2) retrieval, embeddings, and agent memory; (3) orchestration and execution control; (4) interaction contracts between models and tools; and (5) runtime reliability and operational robustness. These families generalize across platform-specific manifestations and capture the core technical difficulties that recur throughout the AI Agent development lifecycle.

To answer \textbf{RQ2}, we measure how popular and difficult different topics are, using established metrics~\cite{uddin2021empirical,alamin2023developer} on Stack Overflow and GitHub issues. On Stack Overflow, developers mostly focus on \emph{Installation \& Dependency Conflicts} and \emph{Prompt \& Output Engineering}. The topics developers find most challenging are \emph{RAG Engineering}, \emph{Document Embeddings \& Vector Stores}, and \emph{Orchestration}, which take longer to resolve and often go unanswered, even though they are less frequently discussed.
GitHub issues reveal additional challenges in real-world use: there are lots of issues about workflows and platforms, but the hardest problems show up in \emph{Agent Orchestration \& Invocation Semantics}, \emph{Policy \& Template Enforcement Friction}, and \emph{Platforms \& Integrations}. On both platforms, configuration and prompting problems are easy to spot and fix, but issues about retrieving information or orchestrating agents are less visible and much harder, highlighting just how complex autonomous, stateful, and tool-based Agent systems can be.
To answer \textbf{RQ3}, we examined GitHub issue data from 18 selected AI Agent repositories. These repositories fall into seven groups based on what they do—for example,\textbf{ core orchestration frameworks, systems that coordinate multiple agents, tools for retrieving information, visual workflow platforms, UI layers, and end-user Agent apps}. We found that each group faces different types of challenges. Infrastructure-focused domains have fewer problems, but those problems are harder to fix and last longer—especially with orchestration and integration. In contrast, UI and workflow platforms see a lot more issues, but these are usually resolved quickly. End-user Agent applications often bring together problems from different areas, showing how decisions made earlier in the architecture and tooling can lead to reliability and operational issues that users experience directly.

The implications of these findings for researchers, practitioners, and educators are discussed. The results provide empirical guidance on prioritizing reliability, observability, and contract enforcement in Agent tooling, documentation, and evaluation methods. 

\section{Methodology}

Stack Overflow provides rich questions, answers, and metadata, but lacks explicit topic annotations for AI agent development. We therefore identify candidate AI-agent posts and categorize them into dominant themes using a multi-step filtering pipeline and LDA topic modeling, a well-established approach for software engineering corpora. To complement Stack Overflow and improve the generalizability of our findings, we also analyze GitHub issues from widely used AI agent frameworks.

\textbf{Step 1: StackExchange Data Explorer for Stack Overflow.}
We collect data using the Stack Exchange Data Explorer\footnote{Stack Exchange Data Explorer. Available at \url{https://data.stackexchange.com}.}, which provides access to official Stack Exchange data dumps \cite{sayyadnejad2024exploring}. We focus on Stack Overflow, as other Stack Exchange sites (e.g., Artificial Intelligence, Data Science, Machine Learning) currently contain limited AI Agent–specific discussions, whereas Stack Overflow offers a large and diverse corpus well-suited for mining practitioner challenges, consistent with prior software engineering studies \cite{elshan2024unveiling, ahmed2025exploring, bi2021mining}.

\textbf{Step 2: Identify AI Agent Tags.}
To identify the most relevant AI Agent-related tags, we followed established tag-expansion strategies from prior studies \cite{abdellatif2020challenges, bagherzadeh2019going, sayyadnejad2024exploring, rosen2016mobile}.
We started querying with the \textit{Agent} tag alone, which returns 1379 questions. To broaden coverage, we expanded our tag set by using the \textit{Agent} co-occurring tags, focusing on avoiding adding noise to our data, since these new tags should help us find AI Agent posts only. We extracted all the \textit{Agent} co-occurring tags from Agent-tagged posts, resulting in 1047 unique co-tags. Next, we applied two heuristic measures to filter and retain only meaningful tags: Tag Relevance Threshold (TRT) and Tag Significance Threshold (TST)~\cite{sayyadnejad2024exploring, rosen2016mobile}. TRT measures how related a specific tag is to the Agent-tagged posts; TST measures how prominent a specific tag is in the Agent-tagged posts.  
We calculate the TRT and TST values as follows~\cite{sayyadnejad2024exploring, rosen2016mobile}:
\begin{equation}
\small
\mathrm{TRT}_{\text{tag}} =
\frac{\text{\# AI Agent posts}}{\text{\# posts}},  \;\;\;\;
\mathrm{TST}_{\text{tag}} =
\frac{\text{\# AI Agent posts}}{\text{\# posts in the initial tag set}}
\end{equation}

A tag was considered both relevant and significant if its corresponding TRT and TST exceeded predetermined thresholds. 
\textcolor{black}{The first two authors}, with expertise in LLM and AI Agent development (particularly for Software Engineering tasks), independently examined the tags under different TRT and TST numerical thresholds. For each tag, we inspected a randomly selected sample of posts to determine when the tags became less relevant to AI Agents; in addition, paying attention to the tag descriptions on Stack Overflow proved necessary and helpful. Ultimately, after testing TRT thresholds \((0.05,\allowbreak 0.10,\allowbreak 0.15,\allowbreak 0.20,\allowbreak 0.25,\allowbreak 0.30)\)
and TST thresholds \((0.001,\allowbreak 0.002,\allowbreak 0.005,\allowbreak 0.010,\allowbreak 0.015,\allowbreak 0.020,\allowbreak 0.30)\), which align closely with previous studies \cite{sayyadnejad2024exploring, rosen2016mobile, abdellatif2020challenges, uddin2021empirical, yang2023understanding}. 
  We also retained tags based on manual validation \cite{sayyadnejad2024exploring, abdellatif2020challenges}. \texttt{langchain-js} and {retrieval-augmented-generation} tag were included because manual verification showed that they consistently refer to AI Agent development, introduce no observable noise, and complement closely related tags that already met the thresholds (e.g., \texttt{langchain}, {langchain-agents}, and \texttt{rag}). This ensured coverage of JavaScript-based agent frameworks and equivalent terminology used for retrieval-augmented agent pipelines.
 This process resulted in a final set of \textbf{15} AI Agent related tags:
{\small
\{\texttt{agent}, \texttt{agents}, \texttt{multi-agent}, \texttt{langgraph},
\texttt{langchain-agents}, \texttt{crewai}, \texttt{ms-autogen},
\texttt{phidata}, {openai-agents}, \texttt{langchain},
\texttt{py-langchain}, \texttt{rag}, \texttt{retrievalqa},
\texttt{langchain-js}, {retrieval-augmented-generation}\}.
}

\textbf{Step 3: Extract AI Agent Posts.}
Using the selected tags, we extracted all Stack Overflow questions (\texttt{PostTypeId}=1) related to AI Agents, yielding 4,783 posts with metadata, consistent with prior work \cite{tahir2018can, sayyadnejad2024exploring}.

\textbf{Step 4 : Data Pre-processing.}
Before conducting topic modeling, we performed several text pre-processing steps to remove noise from the dataset. First, we removed duplicate posts, since searching each tag separately may return the same question multiple times (a post can have multiple tags). This eliminated 740 duplicates. Next, we restricted the dataset to questions posted after the first major AI impact on software development, the release of GitHub Copilot (2021-06-21), which removed 1,172 older posts. We then expanded the dataset by adding accepted answers from Stack Overflow using \texttt{PostTypeId=2}, yielding 530 accepted answers linked to 2,871 questions, for a total of 3,401 pairs of questions and answers. A final validation was conducted to eliminate false positives where the term Agent referred to unrelated concepts, such as Agent-based simulation (e.g., AnyLogic), DevOps and CI/CD execution agents (e.g., Jenkins build agents, GitHub Actions runners, Azure DevOps pipelines), reinforcement-learning agents (e.g., PPO, DQN, Gym-based environments), and robotics or control systems (e.g., ROS-based agents). To derive reliable exclusion criteria, the authors independently inspected random samples of the dataset and then jointly agreed on a consolidated filtering rule through discussion. After applying this rule, the remaining posts were manually re-checked to confirm relevance, resulting in a final dataset of 3,191 AI Agent–related questions with accepted answers. Following established Stack Overflow mining practices \cite{ahmed2018concurrency, barua2014developers, sayyadnejad2024exploring}, code blocks, HTML tags, URLs, and images were removed. A standard natural language processing pipeline was then applied, utilizing Gensim for tokenization and bigrams \cite{vrehuuvrek2010software}, NLTK for stopword removal \cite{loper2002nltk}, and spaCy for lemmatization of informative parts of speech \cite{honnibal2020spacy}. This process yielded a clean, normalized corpus suitable for topic modeling.

\textbf{Step 5: LDA Topic Modeling.} 
%
Latent Dirichlet Allocation (LDA)  is widely used for topic modeling in software repositories~\cite{hindle2016contextual}, including technical Q\&A posts~\cite{sayyadnejad2024exploring, barua2014developers} and issue reports. In this study, we employed the MALLET implementation of LDA~\cite{mccallum2002mallet}, a method widely applied in software engineering research due to its higher coherence scores compared to the Gensim library~\cite{abdellatif2020challenges, li2021understanding, sayyadnejad2024exploring}. LDA leverages two crucial hyperparameters, \(\alpha\) (alpha) and \(\beta\) (beta), which govern the distribution of words across topics and the assignment of posts to topics~\cite{sayyadnejad2024exploring, allamanis2013and, uddin2021empirical, bagherzadeh2019going, ahmed2018concurrency, rosen2016mobile}. We adhered to conventional settings: \(\alpha = 50/K\), where \(K\) is the number of topics, and \(\beta = 0.01\) \cite{bagherzadeh2019going}. These values are widely accepted in prior research and provide a robust benchmark for our experimental framework in topic modeling~\cite{sayyadnejad2024exploring}. Selecting the optimal number of topics (K) is a key challenge in LDA~\cite{panichella2013effectively}, as large K values yield overly specific topics while small K values produce generic ones. We evaluated K values from 5 to 20 using topic coherence, a metric shown to correlate with human judgment~\cite{roder2015exploring}, and repeated LDA–MALLET training three times to reduce stochastic effects. Across all runs, K=7 consistently achieved the highest coherence, with a maximum score of 0.462. Although K=5,6, and 8 showed comparable performance, topics at K=7 were more distinct, a finding independently confirmed by the first two authors through inspection of the top 20 posts per topic. We therefore selected K=7 as the optimal number of topics.


\textbf{Step 6: Sampling.}
 To validate the LDA results while keeping manual analysis feasible, we applied a stratified sampling strategy that accounts for both topic volume and the model’s assignment confidence, following prior empirical studies~\cite{sayyadnejad2024exploring, asgari2025adaptive, uddin2021empirical, ahmed2018concurrency, elshan2024unveiling, abdellatif2020challenges}. We estimated the required sample size using Cochran’s method with a 95\% confidence level and a 5\% margin of error, applying finite population correction due to the bounded corpus size~\cite{woolson1986sample, renthleichoosing, cochran1977sampling}. The resulting sample was proportionally allocated across topics using Neyman’s allocation~\cite{neyman1938contribution, kubiak2022prior}, where within-topic variability was measured using the standard deviation of \texttt{Topic\_Perc\_Contrib}. Within each topic, records were sampled with probability proportional to this confidence score to prioritize high-confidence assignments while preserving topical diversity; the resulting allocation is summarized in Table~\ref{tab:neyman-side-by-side}.

\begin{table}[t]
\centering
\caption{Neyman allocation by topic for Stack Overflow and GitHub Issues.}
\label{tab:neyman-side-by-side}
\scriptsize
\setlength{\tabcolsep}{3pt}
\renewcommand{\arraystretch}{1.05}

\textbf{Stack Overflow}\\
{\footnotesize (Population $N=3{,}191$; $n_0=384.2$; Final Corrected Population $n=343$)}\\[0.35em]
\begin{tabular}{@{}r r r r@{}}
\toprule
\textbf{Topic} & \textbf{Stratum $N_h$} & \textbf{SD $S_h$} & \textbf{Alloc $n_h$} \\
\midrule
1 & 390 & 0.109269 & 41 \\
2 & 554 & 0.116195 & 62 \\
3 & 309 & 0.110661 & 33 \\
4 & 415 & 0.107397 & 43 \\
5 & 667 & 0.115997 & 75 \\
6 & 312 & 0.119681 & 36 \\
7 & 544 & 0.101866 & 53 \\
\midrule
\textbf{Total} & \textbf{3{,}191} &  & \textbf{343} \\
\bottomrule
\end{tabular}

\vspace{0.9em}

\textbf{GitHub Issues}\\
{\footnotesize (Population $N=64{,}098$; $n_0=384.2$; Final Corrected Population $n=382$)}\\[0.35em]
\begin{tabular}{@{}r r r r@{}}
\toprule
\textbf{Topic} & \textbf{Stratum size} & \textbf{Within-SD} & \textbf{Alloc $n_h$} \\
\midrule
1  & 8{,}177 & 0.141207 & 45 \\
2  & 5{,}612 & 0.161634 & 35 \\
3  & 3{,}846 & 0.146409 & 22 \\
4  & 2{,}479 & 0.172917 & 17 \\
5  & 3{,}769 & 0.162406 & 24 \\
6  & 9{,}125 & 0.150321 & 53 \\
7  & 4{,}327 & 0.167707 & 28 \\
8  & 2{,}625 & 0.150413 & 15 \\
9  & 3{,}756 & 0.155045 & 22 \\
10 & 5{,}713 & 0.159515 & 35 \\
11 & 4{,}048 & 0.173298 & 27 \\
12 & 4{,}874 & 0.143373 & 27 \\
13 & 5{,}747 & 0.145429 & 32 \\
\midrule
\textbf{Total} & \textbf{64{,}098} &  & \textbf{382} \\
\bottomrule
\end{tabular}
\end{table}

\textbf{Step 7: Manual Analysis.}

Following prior qualitative topic-interpretation studies \cite{li2021understanding, ahmed2018concurrency, bagherzadeh2019going, alamin2023developer, sayyadnejad2024exploring}, we used a card-sorting approach to label and validate the LDA topics~\cite{fincher2005making, uddin2021empirical}. The first two authors independently reviewed the sampled posts and topic keywords, while the third author audited the process for consistency. After discussion, the authors agreed on topic labels with a Cohen’s Kappa of \(\kappa \approx 0.83\) \cite{mchugh2012interrater}. Using the Neyman-allocated samples (Table\ref{tab:neyman-side-by-side}), the first two authors then iteratively identified and refined sub-topics, converging after multiple rounds of discussion on a final set of 28 sub-topics~\cite{uddin2021empirical, sayyadnejad2024exploring}.

\textbf{Popularity \& Difficulty of Topics.}
We assessed the popularity of AI Agent topics using three standard Stack Overflow metrics widely adopted in prior studies: average view count, average post score, and average number of comments per question~\cite{asgari2024testing, sayyadnejad2024exploring, yang2016security, rosen2016mobile, uddin2021empirical, alamin2023developer, abdellatif2020challenges}. These metrics respectively capture topic visibility, perceived usefulness, and community engagement~\cite{sengupta2020learning, yang2016security}. Topic difficulty was measured using two established proxies: (i) the proportion of questions without accepted answers, and (ii) the median time required to receive an accepted answer~\cite{asgari2024testing, sayyadnejad2024exploring, rosen2016mobile, barua2014developers, bagherzadeh2019going, abdellatif2020challenges}. Higher values on these metrics indicate greater difficulty. To enable comparison across topics, we followed prior work and computed fused popularity and difficulty scores by normalizing each metric relative to the global topic average and then averaging them~\cite{uddin2021empirical, alamin2023developer}. This aggregation provides a concise representation of overall visibility and resolution effort per topic.

\textbf{GitHub Issues Mining \& Topic Modeling.}

To add to our Stack Overflow analysis and make our findings more robust, we also looked at GitHub issues from popular AI Agent repositories. We started with seven core Agent frameworks that came up most often on our Stack Overflow analysis (see Table ~\ref{tab:AIAgentTechnologies-compact}) : \texttt{langchain}, \texttt{langgraph}, \texttt{crewAI}, \texttt{Flowise}, \texttt{llama\_index}, \texttt{semantic-kernel}, and \texttt{autogen}. These frameworks formed the starting point for choosing which repositories to include. Building on earlier software engineering research~\cite{braiek2018open}, we used the number of stars a repository has as a stand-in for how widely it's used and how important it is in the ecosystem. With the official GitHub REST API,\footnote{\url{https://api.github.com}. All queries were executed on December~28,~2025.} we searched for repositories tagged with \emph{Agent} and \emph{Agents} in \textbf{GitHub Topics}, keeping the top five for each topic based on stars. To cover more programming languages than just Python, we also included \texttt{JetBrains/koog} to represent Agent development in Java and Kotlin. In total, we selected 18 repositories (see Table~\ref{tab:repo-metrics}). From these repositories, we gathered 67,193 issues. After removing issues that were empty, one-word, or not in English, we ended up with a final set of 64,098 GitHub issues. We processed this data using the same pipeline and LDA--MALLET setup as in our Stack Overflow analysis. We tested different numbers of topics, from 7 to 40, across three runs. On average, 13 topics ($K=13$) gave us the most consistent and meaningful results, ranking highest in two runs and a close second in the third. To double-check, the first two authors looked at the top 20 issues for each topic in the competing setups ($K=10$ and $K=15$), and agreed that 13 topics made the most sense and were easiest to interpret. For manual validation, we used the same stratified sampling approach as in Step~6, applying Neyman allocation with a correction for finite populations. This gave us a representative sample of 382 GitHub issues (see Table~\ref{tab:neyman-side-by-side}). We then analyzed these issues using the same manual coding process as in Step~7. Agreement between the first two authors was strong (Cohen’s $\kappa = 0.916$), and any disagreements were settled by discussing with a third author.

\section{Results}

\begin{table*}[!tb]
\caption{Stack Overflow Data—Topics and subtopics with within-topic percentage usage.}
\label{tab:taxonomy-panels}
\centering

\begingroup
\scriptsize
\setlength{\tabcolsep}{2pt}
\renewcommand{\arraystretch}{0.95}
\hyphenpenalty=10000
\exhyphenpenalty=10000
\sloppy

\begin{tabularx}{\textwidth}{@{}
  >{\raggedright\arraybackslash}p{.25\textwidth}
  >{\raggedright\arraybackslash}p{.25\textwidth}
  >{\raggedright\arraybackslash}p{.25\textwidth}
  >{\raggedright\arraybackslash}p{.25\textwidth}
@{}}
\toprule
\makecell[l]{\textbf{Topic 1 -- Operational Issues}\\\textbf{(Runtime \& Integration)}\\{\color{gray}\footnotesize\emph{Topic share: 11.95\% of posts}}} &
\makecell[l]{\textbf{Topic 2 -- Document}\\\textbf{Embeddings \& Vector Stores}\\{\color{gray}\footnotesize\emph{Topic share: 18.08\% of posts}}} &
\makecell[l]{\textbf{Topic 3 -- Robustness,}\\\textbf{Reliability \& Evaluation}\\{\color{gray}\footnotesize\emph{Topic share: 9.62\% of posts}}} &
\makecell[l]{\textbf{Topic 4 -- Orchestration \&}\\\textbf{Workflow Control}\\{\color{gray}\footnotesize\emph{Topic share: 12.54\% of posts}}} \\
\midrule

\begin{tabular}{@{}>{\raggedright\arraybackslash}p{.78\linewidth}r@{}}
\textbf{Subtopic} & \textbf{\%}\\ \midrule
Packaging \& Environments (Build/CI/CD) & 17.07\\
Model Runtimes \& Performance & 12.20\\
Connectivity \& Cloud Operations & 24.39\\
Application/SDK Integration \& Data Pipelines & 46.34\\
\end{tabular}
&
\begin{tabular}{@{}>{\raggedright\arraybackslash}p{.78\linewidth}r@{}}
\textbf{Subtopic} & \textbf{\%}\\ \midrule
Ingestion \& Context Modeling & 30.65\\
Vector Storage \& Index Management & 25.81\\
Retrieval Quality \& Similarity Signals & 14.52\\
Store Integrations \& Runtime Ops & 29.03\\
\end{tabular}
&
\begin{tabular}{@{}>{\raggedright\arraybackslash}p{.78\linewidth}r@{}}
\textbf{Subtopic} & \textbf{\%}\\ \midrule
Evaluation \& Reproducibility & 9.09\\
Learning \& Planning Correctness & 15.15\\
Contracts, Tools \& Grounding & 42.42\\
Operational Reliability \& Instrumentation & 33.33\\
\end{tabular}
&
\begin{tabular}{@{}>{\raggedright\arraybackslash}p{.78\linewidth}r@{}}
\textbf{Subtopic} & \textbf{\%}\\ \midrule
Graph \& State Architecture & 26.83\\
Execution Policies \& Human Control & 31.71\\
Persistence, Concurrency \& Real-Time I/O & 7.32\\
Integration, Services \& Observability & 34.15\\
\end{tabular}
\\
\midrule
\end{tabularx}

\vspace{0.6em}

\begin{tabularx}{\textwidth}{@{}
  >{\raggedright\arraybackslash}p{.333\textwidth}
  >{\raggedright\arraybackslash}p{.333\textwidth}
  >{\raggedright\arraybackslash}p{.333\textwidth}
@{}}
\makecell[l]{\textbf{Topic 5 -- Installation \&}\\\textbf{Dependency Conflicts}\\{\color{gray}\footnotesize\emph{Topic share: 21.90\% of posts}}} &
\makecell[l]{\textbf{Topic 6 -- RAG Engineering}\\{\color{gray}\footnotesize\emph{Topic share: 10.50\% of posts}}} &
\makecell[l]{\textbf{Topic 7 -- Prompt \& Output}\\\textbf{Engineering}\\{\color{gray}\footnotesize\emph{Topic share: 15.45\% of posts}}} \\
\midrule

\begin{tabular}{@{}>{\raggedright\arraybackslash}p{.78\linewidth}r@{}}
\textbf{Subtopic} & \textbf{\%}\\ \midrule
Python Ecosystem Dependency Health & 20.00\\
Framework \& SDK API Churn & 49.33\\
Platform Compatibility Issues & 18.67\\
Provider \& Vector Service Compatibility Issues & 12.00\\
\end{tabular}
&
\begin{tabular}{@{}>{\raggedright\arraybackslash}p{.78\linewidth}r@{}}
\textbf{Subtopic} & \textbf{\%}\\ \midrule
RAG Ingestion \& Retrieval Modeling & 22.22\\
RAG Indexes, Stores \& Caching & 13.89\\
RAG Scale, Cost \& Runtime Operations & 30.56\\
RAG Design, Evaluation \& Query Strategy & 33.33\\
\end{tabular}
&
\begin{tabular}{@{}>{\raggedright\arraybackslash}p{.78\linewidth}r@{}}
\textbf{Subtopic} & \textbf{\%}\\ \midrule
Prompt Design \& Chat Orchestration & 56.60\\
JSON \& Tool-Calling Output Contracts & 18.87\\
Memory \& Response Shaping & 22.64\\
Multimodal Prompting \& Output Control & 1.89\\
\end{tabular}
\\
\bottomrule
\end{tabularx}

\endgroup
\end{table*}

\begin{table*}[!tb]
\caption{GitHub Issues Data—Topic Taxonomy.}
\label{tab:gitdata-taxonomy}
\centering\footnotesize
\setlength{\tabcolsep}{6pt}
\renewcommand{\arraystretch}{1.12}

\begin{tabularx}{\textwidth}{@{}l Y@{}}
\toprule
\textbf{Topic} & \textbf{Scope} \\
\midrule
\textbf{Topic 1: Platforms \& Integrations} {\color{gray}\emph{(Stratum: 12.76\%)}} &
Cross-platform support, SDKs, connectors, cloud services, and external system integrations. \\

\textbf{Topic 2: Embeddings \& Retrieval} {\color{gray}\emph{(Stratum: 8.76\%)}} &
Vector databases, embedding generation, indexing, similarity search, and RAG retrieval pipelines. \\

\textbf{Topic 3: Model--Agent--Tool Interaction Contracts (Prompting \& Parsing)} {\color{gray}\emph{(Stratum: 6.00\%)}} &
Failures to follow structured interaction contracts, including prompt formats, tool calling schemas,
action inputs, multimodal inputs, and output parsing. \\

\textbf{Topic 4: Version-Update Regressions} {\color{gray}\emph{(Stratum: 3.87\%)}} &
Bugs and behavior changes introduced by library, framework, or model version upgrades. \\

\textbf{Topic 5: Dependencies, Installs, Imports, Environments} {\color{gray}\emph{(Stratum: 5.88\%)}} &
Package installation issues, dependency conflicts, import errors, environment setup, and version pinning. \\

\textbf{Topic 6: Workflow, UI \& API Behavior} {\color{gray}\emph{(Stratum: 14.24\%)}} &
Workflow logic, user interface behavior, REST/SDK API usage, request/response handling, and UX issues. \\

\textbf{Topic 7: Feature Requests \& Configurability} {\color{gray}\emph{(Stratum: 6.75\%)}} &
Requests for new features, enhancements, configuration options, and extensibility. \\

\textbf{Topic 8: Tracebacks \& Unhandled Exceptions} {\color{gray}\emph{(Stratum: 4.10\%)}} &
Crashes, stack traces, uncaught exceptions, and errors directly surfaced to users or logs. \\

\textbf{Topic 9: LangChain Runtime \& Integration Breakages} {\color{gray}\emph{(Stratum: 5.86\%)}} &
Runtime failures, breaking changes, and integration issues specific to the LangChain ecosystem. \\

\textbf{Topic 10: Policy \& Template Enforcement Friction} {\color{gray}\emph{(Stratum: 8.91\%)}} &
Prompt policies, guardrails, templates, schema enforcement, validation constraints, and compliance friction. \\

\textbf{Topic 11: Runtime Execution \& Operational Failures (Docker, CLI, Headless)} {\color{gray}\emph{(Stratum: 6.32\%)}} &
Execution failures in deployment and operations, including Docker, CLI tools, servers, and headless runs. \\

\textbf{Topic 12: Agent Orchestration \& Invocation Semantics} {\color{gray}\emph{(Stratum: 7.60\%)}} &
Multi-agent coordination, sequencing, routing, invocation order, concurrency, and orchestration logic. \\

\textbf{Topic 13: Provider / Model Inconsistencies \& Tooling Limits} {\color{gray}\emph{(Stratum: 8.97\%)}} &
Provider- or model-specific behavior differences, capability gaps, API mismatches, and tooling limitations. \\
\bottomrule
\end{tabularx}
\end{table*}

\subsection{RQ1: Taxonomy of Challenges}

To address RQ1, complementary topic taxonomies were constructed from Stack Overflow questions and GitHub issues related to AI Agent development. Both platforms capture developer-reported problems: Stack Overflow questions describe difficulties, while GitHub issues document failures, limitations, or unexpected behavior. The extracted topics are therefore interpreted as recurring technical challenges encountered by Agent developers. At the first level, fine-grained, topic taxonomies were derived. Stack Overflow yielded seven topics comprising 28 subtopics (ST; Table~\ref{tab:taxonomy-panels}), reflecting common difficulties related to usage and understanding. GitHub yielded thirteen issue topics (GT; Table~\ref{tab:gitdata-taxonomy}) that capture recurring implementation, integration, and maintenance problems reported across Agent frameworks. For clearer interpretation and systematic comparison across platforms, these fine-grained topics were further organized into \textbf{five higher-level families of major Agent challenges}: dependency and environment management, retrieval and memory pipelines, orchestration and execution control, interaction contracts, and runtime reliability. These families form a hierarchical abstraction that groups related Stack Overflow (ST) and GitHub (GT) topics by their shared underlying technical difficulty rather than their platform-specific surface manifestations. The five challenge families were derived through open coding and consensus-based synthesis over the extracted topics. Two authors independently examined topic titles, top keywords, and a stratified sample of 15 representative Stack Overflow posts or GitHub issues per topic to identify the core difficulty each topic represents. Based on this material, the authors independently grouped ST and GT topics into higher-level challenge families, focusing on root technical concerns rather than platform-specific phrasing. Inter-rater agreement was assessed using Cohen’s $\kappa$ = 0.825, indicating substantial agreement. Remaining disagreements were resolved through discussion with a third author to ensure conceptual consistency. Across both data sources, the most prominent challenge families consistently involve environment and dependency management (ST5; GT1, GT4, GT5), retrieval and knowledge integration within RAG pipelines (ST2, ST6; GT2, GT13), and Agent orchestration and execution control (ST4; GT12). Interaction contracts, which cover prompting, tool invocation, and structured outputs, form another major family (ST7; GT3, GT10), while runtime reliability and operational robustness issues frequently arise during execution and deployment (ST1, ST3; GT8, GT9, GT11). Collectively, these families provide a unified, hierarchical view of the technical challenges that characterize contemporary AI Agent development.

\textbf{Major Challenge 1: Environment, Platforms, and Dependency Management.}
This challenge concerns configuring execution environments, resolving dependency conflicts, and integrating Agent frameworks with external platforms and services. It is Agent-specific because Agents combine multiple tightly coupled components—models, tools, memory stores, runtimes, and UI or service layers—where version drift or platform mismatches can break end-to-end execution. On Stack Overflow, this challenge appears primarily as \emph{Installation \& Dependency Conflicts} (\textbf{ST5}), where developers report failures caused by framework API churn or incompatible library upgrades (e.g., \href{https://stackoverflow.com/q/77338572}{Q77338572}, \href{https://stackoverflow.com/q/78613825}{Q78613825}).  In contrast, GitHub issues emphasize platform-level and integration failures encountered during real deployments, including \emph{Platforms \& Integrations} (\textbf{GT1}), \emph{Dependencies, Installs, Imports, Environments} (\textbf{GT5}), and \emph{Version-Update Regressions} (\textbf{GT4}). Representative examples include cross-platform runtime and SDK integration failures in OpenHands (e.g., \href{https://github.com/OpenHands/OpenHands/issues/11521}{I11521}), environment and import errors preventing Agent initialization in AutoGen (e.g., \href{https://github.com/microsoft/autogen/issues/2756}{I2756}), and breaking changes introduced by LangChain upgrades that invalidate existing Agent pipelines (e.g., \href{https://github.com/langchain-ai/langchain/issues/23517}{I23517}). This contrast highlights how Stack Overflow reflects setup-time confusion, while GitHub exposes deployment-time fragility and integration risk in production Agent systems.

\textbf{Major Challenge 2: Retrieval, Embeddings, and Agent Memory.}
AI Agents rely on retrieval pipelines and persistent memory to ground decisions across multiple steps and interactions, making retrieval quality and state management central concerns. This challenge goes beyond general LLM applications, which typically retrieve context once per request, whereas Agents repeatedly query, update, and reason over stored knowledge. On Stack Overflow, these difficulties surface primarily as \emph{Document Embeddings \& Vector Stores} and \emph{RAG Engineering} (\textbf{ST2}, \textbf{ST6}), where developers struggle with document chunking strategies, indexing semantics, and performance bottlenecks in vector stores (e.g., \href{https://stackoverflow.com/q/78015622}{Q78015622}, \href{https://stackoverflow.com/q/78505822}{Q78505822}). These discussions reflect design-time uncertainty about how to structure and persist Agent memory. In contrast, GitHub issues emphasize deployment-time and framework-level failures captured by \emph{Embeddings \& Retrieval} (\textbf{GT2}) and \emph{Provider / Model Inconsistencies \& Tooling Limits} (\textbf{GT13}). Representative issues include embedding and retrieval mismatches in LlamaIndex that break Agent memory pipelines (e.g., \href{https://github.com/run-llama/llama_index/issues/13488}{I13488}) and provider-specific behavior that causes inconsistent retrieval results across Agent runs (e.g., \href{https://github.com/langflow-ai/langflow/issues/5980}{I5980}).  Together, these findings show that retrieval in Agent systems is not merely a modeling concern but a core operational dependency that directly affects planning correctness, tool invocation, and long-horizon Agent behavior.

\textbf{Major Challenge 3: Orchestration and Execution Control.}
This challenge captures how Agents coordinate tools, regulate execution flow, and maintain control across multi-step or multi-Agent workflows. Orchestration is inherently Agent-specific because it governs delegation, sequencing, and execution boundaries beyond a single LLM invocation. On Stack Overflow, this challenge appears as \emph{Orchestration \& Workflow Control} (\textbf{ST4}), where developers raise questions about controlling tool invocation policies and diagnosing missing or ambiguous observability in complex multi-Agent workflows (e.g., \href{https://stackoverflow.com/q/79332599}{Q79332599}, \href{https://stackoverflow.com/q/79771469}{Q79771469}). In contrast, GitHub issues reveal orchestration breakdowns that emerge during real-world execution, captured by \emph{Agent Orchestration \& Invocation Semantics} (\textbf{GT12}). For example, Semantic Kernel reports failures where Agents surface function-call intents during streaming but do not execute them, leading to stalled execution despite correct tool configuration (e.g., \href{https://github.com/microsoft/semantic-kernel/issues/9245}{I9245}). This contrast highlights how Stack Overflow surfaces design-time uncertainty and observability gaps, while GitHub exposes execution-time orchestration failures in deployed Agent systems.

\textbf{Major Challenge 4: Interaction Contracts Between Models and Tools.}
AI Agents depend on strict interaction contracts governing prompts, tool calls, parameter schemas, and structured outputs. These contracts are critical because even minor violations can halt planning loops, break execution, or cause silent failures that are difficult to diagnose. On Stack Overflow, such issues are discussed under \emph{Prompt \& Output Engineering} (\textbf{ST7}), where developers report difficulties enforcing JSON schemas, constraining model outputs, and reliably triggering tool calls (e.g., \href{https://stackoverflow.com/q/77227902}{Q77227902}, \href{https://stackoverflow.com/q/77362103}{Q77362103}). These discussions emphasize prompt design and output validation as mechanisms for maintaining contract compliance at development time. GitHub issues reveal the downstream consequences of contract violations in deployed systems, captured by \emph{Model–Agent–Tool Interaction Contracts} (\textbf{GT3}). For example, in RagFlow, a framework upgrade removed an expected tool-side parameter class, causing agents to fail at runtime due to broken assumptions about tool interfaces (e.g., \href{https://github.com/infiniflow/ragflow/issues/9141}{I9141}). Similarly, LlamaIndex users report the inability to constrain generated entity and relation types when constructing knowledge graphs, highlighting the absence of enforceable schemas between model outputs and downstream tool expectations (e.g., \href{https://github.com/run-llama/llama_index/issues/10120}{I10120}). Together, these issues show how interaction contracts evolve from a prompt-level usability concern on Stack Overflow into a correctness, robustness, and versioning challenge in real-world Agent systems.

\textbf{Major Challenge 5: Runtime Reliability and Operational Robustness.}
The final challenge concerns keeping AI Agents reliable during execution, deployment, and long-running operation. Agents are particularly vulnerable because they execute over extended periods, invoke external services, and maintain evolving state. On Stack Overflow, this challenge appears in \emph{Operational Issues (Runtime \& Integration)} and \emph{Robustness, Reliability \& Evaluation} (\textbf{ST1}, \textbf{ST3}), for example when permission or configuration errors invalidate evaluations (e.g., \href{https://stackoverflow.com/q/76815315}{Q76815315}) or malformed tool outputs repeatedly break execution (e.g., \href{https://stackoverflow.com/q/79134666}{Q79134666}). GitHub issues emphasize the operational consequences of these problems, captured by \emph{Tracebacks \& Unhandled Exceptions} (\textbf{GT8}), \emph{Runtime Execution \& Operational Failures} (\textbf{GT11}), and ecosystem-specific breakages such as LangChain runtime failures (\textbf{GT9}). Representative examples include repeated crashes in MetaGPT deployments (e.g., \href{https://github.com/FoundationAgents/MetaGPT/issues/1862}{I1862}), execution failures in OpenHands during long-running tasks (e.g., \href{https://github.com/OpenHands/OpenHands/issues/907}{I907}), and unhandled runtime errors in LangChain Agent pipelines (e.g., \href{https://github.com/langchain-ai/langchain/issues/7900}{I7900}).

\textbf{Workflow, UI \& API Behavior} (\textbf{GT6}) constitutes the largest stratum in the GitHub Issues dataset (14.24\%), underscoring its importance in practice. These issues primarily fall under Challenge 3 and Challenge 1, as they concern how Agent workflows are composed, triggered, and executed through user interfaces and APIs. Most cases arise from UI-centric Agent platforms—particularly langgenius/dify—where Agents are built and deployed via visual workflows rather than code alone (e.g., \href{https://github.com/langgenius/dify/issues/15785}{I15785}). This reflects a shift toward production-grade, user-facing Agent systems, where orchestration complexity and integration failures increasingly surface at the workflow and interface layer.

\textbf{Feature Requests \& Configurability} (\textbf{GT7}) covers a unique but related set of Agent challenges, making up 6.75\% of GitHub issues. Rather than focusing on bugs or failures, these issues show that developers often need more ways to extend, customize, or control how Agents work and fit into their infrastructure. Many of these requests come up after Agents are used in real-world settings, where the built-in features no longer meet all needs. For example, developers might want more configuration options, better API access, or easier ways to connect Agents with existing company systems. One case is developers using langgenius/dify who asked for extra OpenAPI endpoints so they could manage Agent annotations through code, pointing to a gap between what the UI can do and what automation requires (see \href{https://github.com/langgenius/dify/issues/6878}{I6878}). Most of these issues relate to Challenge~1 (integration with environments and platforms) and Challenge~3 (control over orchestration and execution), since they are about making Agent systems more adaptable and manageable beyond default setups. The rise of GT7 shows that as Agent platforms get more advanced, developers spend less time fixing basic errors and more time seeking flexibility, compatibility, and maintainability for the long term.

\begin{center}
\fbox{\parbox{0.98\linewidth}{
\textbf{Findings (AI Agent Topics).}
AI Agent development involves a distinct and recurring set of technical challenges that cannot be fully explained by general LLM-based application development. These challenges span environment setup and dependency management, retrieval and memory pipelines, multi-agent orchestration, model–tool interaction contracts, runtime reliability, and workflow- and configuration-level concerns. Their recurrence indicates that Agent-specific difficulty arises not only from autonomy, tool-mediated execution, and long-running behavior, but also from the need to compose, customize, and govern Agents through user interfaces, APIs, and extensible configurations rather than isolated framework flaws or misuse.
}}
\end{center}

\subsection{RQ2: Popularity \& Difficulty}
To address RQ2, we analyze the relative popularity and difficulty of AI Agent challenges using complementary signals from Stack Overflow and GitHub issues.

\textbf{Popularity of Agent Challenges (Stack Overflow).}
Developer attention on Stack Overflow is concentrated around a small set of recurring topics (Table~\ref{tab:popdiff-fused}). The most popular challenges are \emph{Installation \& Dependency Conflicts} (\textbf{ST5}), followed by \emph{Prompt \& Output Engineering} (\textbf{ST7}) and \emph{Document Embeddings \& Vector Stores} (\textbf{ST2}). These topics reflect common entry points into Agent development, where practitioners seek help configuring environments, controlling Agent behavior, and enabling retrieval-based memory. In contrast, \emph{Orchestration} (\textbf{ST4}) and \emph{RAG Engineering} (\textbf{ST6}) receive comparatively less attention, suggesting that advanced Agent workflows affect a smaller, more specialized subset of developers.

\textbf{Difficulty of Agent Challenges (Stack Overflow).}
Difficulty on Stack Overflow follows a different pattern. The hardest topics are \emph{RAG Engineering} (\textbf{ST6}), \emph{Document Embeddings \& Vector Stores} (\textbf{ST2}), and \emph{Orchestration} (\textbf{ST4}), which exhibit long resolution times and a high share of unanswered questions. These challenges involve multi-step reasoning, persistent state, and coordination across tools, capabilities that are intrinsic to Agent systems. By contrast, although \emph{Installation \& Dependency Conflicts} (\textbf{ST5}) is the most frequently discussed topic, it is among the easiest to resolve, indicating that such issues are widespread but relatively well understood once the root cause is identified.

\textbf{Issue Volume and Resolution Difficulty (GitHub).}
GitHub issues capture a different perspective, reflecting engineering workload and maintenance burden in production Agent frameworks. The largest issue volumes are observed for \emph{Workflow, UI \& API Behavior} (\textbf{GT6}), \emph{Platforms \& Integrations} (\textbf{GT1}), and \emph{Policy \& Template Enforcement Friction} (\textbf{GT10}), highlighting the operational challenges of deploying user-facing, platform-integrated Agents. Resolution difficulty is most pronounced for \emph{Policy \& Template Enforcement Friction} (\textbf{GT10}), \emph{Agent Orchestration \& Invocation Semantics} (\textbf{GT12}), and \emph{Platforms \& Integrations} (\textbf{GT1}), which exhibit high open-issue rates and prolonged activity. In contrast, topics such as \emph{Workflow, UI \& API Behavior} (\textbf{GT6}) and \emph{Feature Requests \& Configurability} (\textbf{GT7}) tend to close quickly, indicating that while frequent, these issues are often localized and easier for maintainers to address.

\textbf{Stack Overflow vs. GitHub Comparison.}
Taken together, the two platforms reveal complementary views of Agent difficulty. Stack Overflow highlights where developers struggle conceptually and during early development, with retrieval, orchestration, and RAG emerging as the hardest topics despite limited visibility. GitHub, in contrast, exposes where Agent systems fail at scale: orchestration semantics, policy enforcement, and platform integration translate into sustained maintenance effort and unresolved issues. Notably, \emph{Orchestration} shifts from being niche-but-hard on Stack Overflow (\textbf{ST4}) to a mid-to-high volume and difficult topic on GitHub (\textbf{GT12}), reflecting its growing importance as Agents move from prototypes to production systems.

\begin{figure}
    \centering
    \includegraphics[width=1\linewidth]{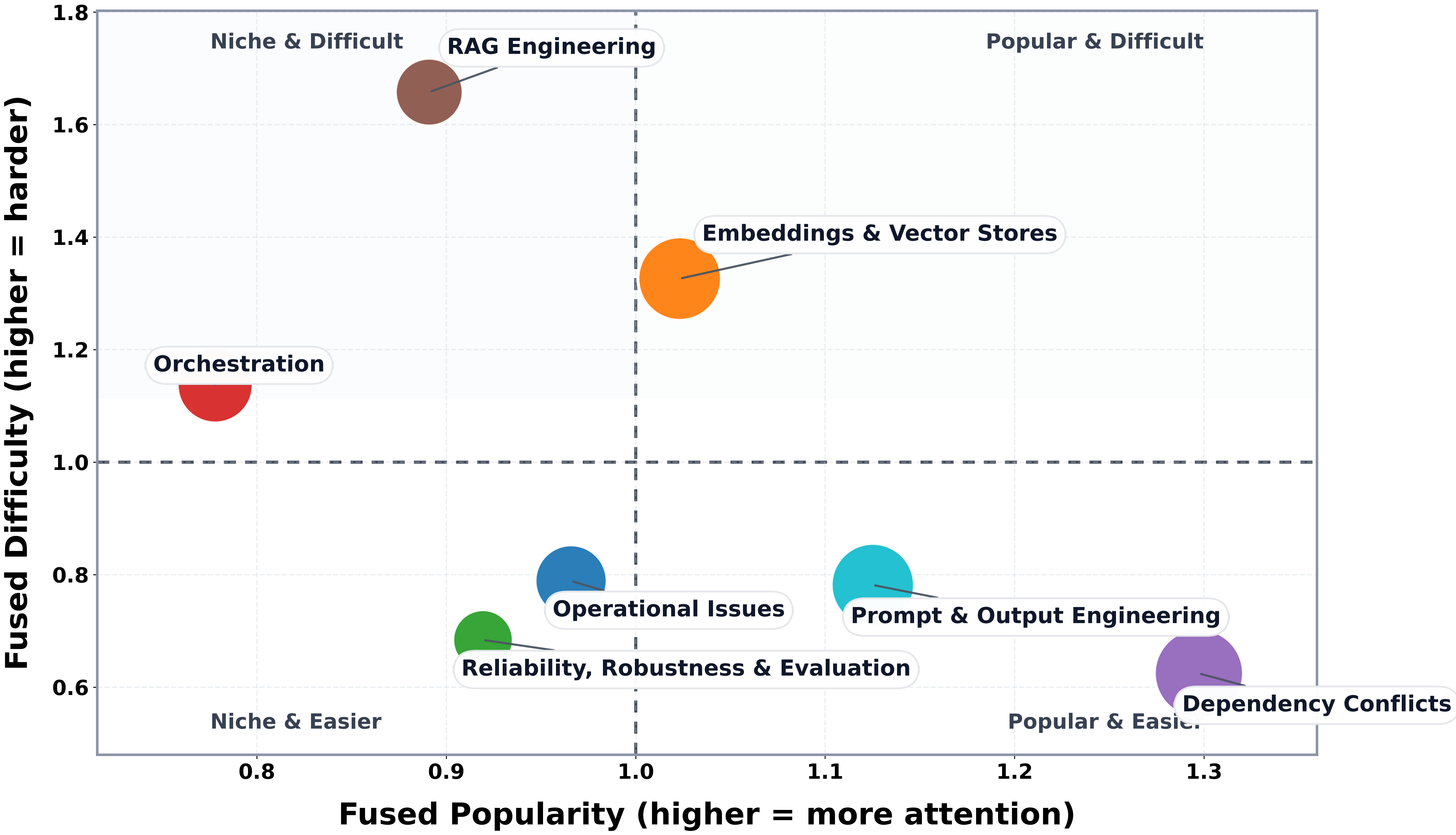}
    \caption{Stack Overflow Fused Popularity \& Fused Difficulty.}
    \label{fig:FPopularity-FDifficulty}
\end{figure}

\begin{figure}
    \centering
    \includegraphics[width=1\linewidth]{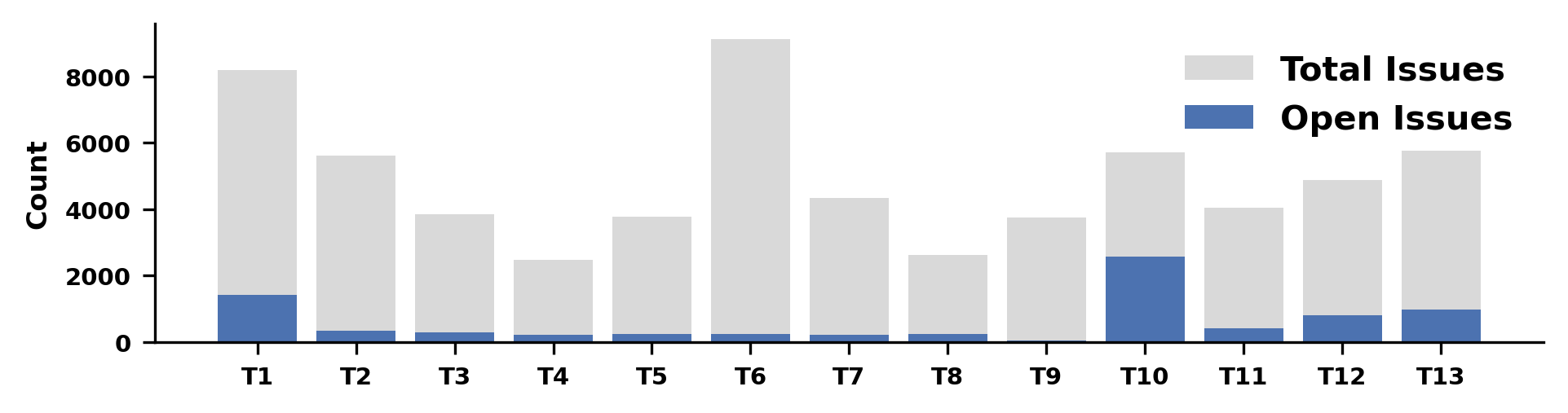}
    \caption{GitHub Issues Topics.}
    \label{fig:issues_open_vs_total_by_topic}
\end{figure}

\begin{center}
\fbox{\parbox{0.98\linewidth}{
\textbf{Findings (Popularity \& Difficulty).}
Easy-to-fix configuration and prompting issues dominate attention, while retrieval- and orchestration-centric challenges are less visible but substantially harder. As Agent systems transition into production, these harder challenges reappear on GitHub as prolonged operational and maintenance burdens, underscoring that Agent difficulty stems from autonomy, statefulness, and tool-mediated execution rather than isolated implementation errors.
}}
\end{center}

\begin{table*}[t]
\caption{Popularity and difficulty of AI Agent topics.}
\label{tab:popdiff-fused}
\centering
\begingroup
\footnotesize
\setlength{\tabcolsep}{3pt}
\renewcommand{\arraystretch}{0.95}
\begin{tabularx}{\textwidth}{@{}l
  >{\raggedleft\arraybackslash}X
  >{\raggedleft\arraybackslash}X
  >{\raggedleft\arraybackslash}X
  >{\raggedleft\arraybackslash}X
  >{\raggedleft\arraybackslash}X
  >{\raggedleft\arraybackslash}X
  >{\raggedleft\arraybackslash}X
  >{\raggedleft\arraybackslash}X@{}}
\toprule
\textbf{Topic} &
\textbf{Views (avg)} &
\textbf{Comments (avg)} &
\textbf{Answers (avg)} &
\textbf{Score (avg)} &
\textbf{No accepted ans.\ (share)} &
\textbf{Median hrs.\ to accepted} &
\textbf{Fused Popularity} &
\textbf{Fused Difficulty} \\
\midrule
1.\ Operations (Runtime \& Integration)                & 1759.70 & 1.082 & 0.915 & 0.848 & 0.827 & 22.52 & \textbf{0.97} & \textbf{0.79} \\
2.\ Document Embeddings \& Vector Stores      & 2426.09 & 0.626 & 0.965 & 1.265 & 0.805 & 65.38 & \textbf{1.02} & \textbf{1.33} \\
3.\ Robustness, Reliability \& Evaluation     & 1441.01 & 1.031 & 0.855 & 0.953 & 0.772 & 16.92 & \textbf{0.92} & \textbf{0.68} \\
4.\ Orchestration         & 1259.11 & 0.768 & 0.734 & 0.905 & 0.884 & 46.93 & \textbf{0.78} & \textbf{1.14} \\
5.\ Installation \& Dependency Conflicts      & 3231.54 & 1.166 & 1.096 & 1.200 & 0.785 & 11.71 & \textbf{1.30} & \textbf{0.62} \\
6.\ RAG Engineering                   & 1677.71 & 0.678 & 0.746 & 1.286 & 0.884 & 87.44 & \textbf{0.89} & \textbf{1.66} \\
7.\ Prompt \& Output Engineering      & 2621.59 & 0.778 & 0.975 & 1.409 & 0.836 & 21.51 & \textbf{1.13} & \textbf{0.78} \\
\midrule
\textit{All topics (avg)}                     & \textbf{2059.54} & \textbf{0.876} & \textbf{0.898} & \textbf{1.124} & \textbf{0.828} & \textbf{38.92} & \textbf{1.00} & \textbf{1.00} \\
\bottomrule
\end{tabularx}
\endgroup
\end{table*}

\begin{table}[t]
\caption{Issue metrics by topic (medians). Close d = median time-to-close (days, closed issues only). Open d = median age (days, open issues only). Act h = median activity span (hours).}
\centering
\scriptsize
\setlength{\tabcolsep}{3pt}
\renewcommand{\arraystretch}{0.9}
\begin{tabular}{lrrrrr}
\toprule
Topic & Open\% & Close d & Open d & Act h & Com \\
\midrule
T1  & 17.4 & 39.9 & 214.2 & 1064.6 & 2 \\
T2  & 6.2  & 35.4 & 217.0 & 1304.7 & 3 \\
T3  & 7.7  & 50.4 & 310.6 & 1647.4 & 3 \\
T4  & 8.7  & 32.4 & 133.3 & 1331.1 & 2 \\
T5  & 6.5  & 6.0  & 211.4 & 300.2  & 3 \\
T6  & 2.7  & 0.4  & 48.9  & 24.9   & 2 \\
T7  & 4.9  & 1.8  & 125.1 & 115.6  & 1 \\
T8  & 9.2  & 4.1  & 247.7 & 1352.5 & 4 \\
T9  & 0.9  & 98.4 & 185.3 & 2382.4 & 3 \\
T10 & 44.9 & 2.2  & 253.8 & 89.1   & 3 \\
T11 & 10.3 & 6.7  & 270.1 & 342.6  & 3 \\
T12 & 16.3 & 15.0 & 161.5 & 550.6  & 2 \\
T13 & 17.2 & 11.6 & 244.8 & 893.4  & 3 \\

\bottomrule
\end{tabular}

\label{tab:topic_metrics_tiny}
\end{table}

\subsection{RQ3: Domains of the AI Agent Ecosystem}

To address RQ3, we organized the GitHub repositories from Table~\ref{tab:repo-metrics} into seven ecosystem domains according to their functional roles and intended uses within the AI Agent landscape. These domains illustrate the positioning of Agent technologies in the ecosystem, spanning core orchestration frameworks, memory components, visual workflow platforms, and end-user Agent applications. We assigned domains through a collaborative, consensus-driven qualitative analysis of the repositories listed in Table~\ref{tab:repo-metrics}. Three authors jointly examined each repository’s README file, description, and, when available, official project documentation or website to determine its primary purpose, target users, and architectural role. Additionally, we reviewed a random sample of 15 GitHub issues per repository to identify typical usage scenarios and failure modes. Drawing on this material, we iteratively discussed and refined domain boundaries until reaching agreement for all repositories. We selected this consensus-based approach because ecosystem domains are interpretive constructs that depend on repository intent and usage context rather than isolated technical features. The resulting seven domains represent a structured synthesis grounded in repository documentation, observed issue patterns, and shared expert judgment.

\textbf{UI / Copilot Components} focus on embedding Agents into interactive user experiences. Although represented by a single primary repository (\texttt{CopilotKit}, Table~\ref{tab:repo-metrics}), this domain shows a high concentration of issues related to \emph{Workflow, UI \& API Behavior} (GT6) and \emph{Runtime Execution \& Operational Failures} (GT11). This indicates that even thin UI layers amplify reliability and integration problems once Agents operate in real-time, user-facing settings.

\textbf{Memory \& State Components}, represented primarily by \texttt{mem0}, are dominated by \emph{Embeddings \& Retrieval} (GT2) and \emph{Provider / Model Inconsistencies} (GT13). These challenges stem from maintaining persistent Agent memory across providers and execution contexts, and they tend to persist longer due to the architectural nature of state management.

\textbf{Multi-Agent Coordination Frameworks}, including \texttt{AutoGen}, \texttt{CrewAI}, and \texttt{MetaGPT}, exhibit heavy concentration in \emph{Agent Orchestration \& Invocation Semantics} (GT12) and \emph{Dependencies, Installs, Imports, Environments} (GT5), with several topics exceeding one thousand issues. Delegation, shared memory, and inter-Agent communication amplify failure modes that are difficult to localize and resolve.

\textbf{Core Orchestration Frameworks} (e.g., \texttt{langchain}, \texttt{langgraph}, \texttt{semantic-kernel}) form the backbone of the ecosystem and account for some of the largest issue volumes overall. Their dominant challenges include \emph{Platforms \& Integrations} (GT1), \emph{Version-Update Regressions} (GT4), and \emph{Agent Orchestration \& Invocation Semantics} (GT12). These issues exhibit long activity spans, highlighting the fragility introduced by API churn in foundational Agent infrastructure.

\textbf{RAG \& Knowledge Systems}, led by \texttt{ragflow} and \texttt{llama\_index}, are heavily dominated by \emph{Embeddings \& Retrieval} (GT2) and \emph{Policy \& Template Enforcement Friction} (GT10), each with several thousand issues. This domain underscores that retrieval quality, grounding guarantees, and policy enforcement remain persistent technical barriers for knowledge-intensive Agents.

\textbf{Visual Workflow Platforms} constitute the largest domain by issue volume, driven primarily by \texttt{langgenius/dify}, \texttt{langflow}, and \texttt{Flowise}. Challenges are overwhelmingly concentrated in \emph{Workflow, UI \& API Behavior} (GT6) and \emph{Feature Requests \& Configurability} (GT7). These issues tend to resolve quickly, suggesting that low-code Agent builders generate frequent but localized problems as users assemble production workflows through graphical abstractions.

\textbf{Agent Apps (End-User)} aggregate failures across the entire stack. Repositories such as \texttt{lobe-chat} and \texttt{OpenHands} each report well over one thousand issues, spanning \emph{Platforms \& Integrations} (GT1), \emph{Runtime Execution \& Operational Failures} (GT11), \emph{Provider / Model Inconsistencies} (GT13), and \emph{Tracebacks \& Unhandled Exceptions} (GT8). This domain exposes how upstream design decisions manifest as user-visible failures when Agents are deployed as full applications.

\paragraph{Domain-Level Difficulty and Resolution Dynamics.}
Table~\ref{tab:cat_metrics_tiny} highlights systematic differences in how challenges evolve across Agent ecosystem domains. \emph{UI / Copilot Components} exhibit the highest open-issue rate, reflecting the difficulty of stabilizing interactive Agent behavior amid rapid UI iteration. \emph{Core Orchestration Frameworks} show the longest activity spans and time-to-close, consistent with the architectural depth of orchestration logic and the need for careful regression control. In contrast, \emph{Visual Workflow Platforms} and \emph{RAG \& Knowledge Systems} tend to close issues quickly, indicating that many reports are incremental or configuration-driven. \emph{Agent Apps (End-User)} combine moderate open rates with long activity spans, suggesting that application-level Agents surface compounded failures originating across multiple upstream domains. Overall, Agent difficulty varies systematically by domain role rather than uniformly across the ecosystem.

\begin{center}
\fbox{\parbox{0.98\linewidth}{
\textbf{Findings (Domains).}
The AI Agent ecosystem is structurally heterogeneous: infrastructure-oriented domains concentrate fewer but harder orchestration and integration challenges, while UI- and workflow-oriented domains generate large volumes of faster-resolving issues. End-user Agent applications amplify challenges across all domains, confirming that Agent-specific difficulty arises from autonomy, persistent state, and cross-system coordination rather than isolated framework defects.
}}
\end{center}

\begin{figure*}
    \centering
    \includegraphics[width=1\linewidth]{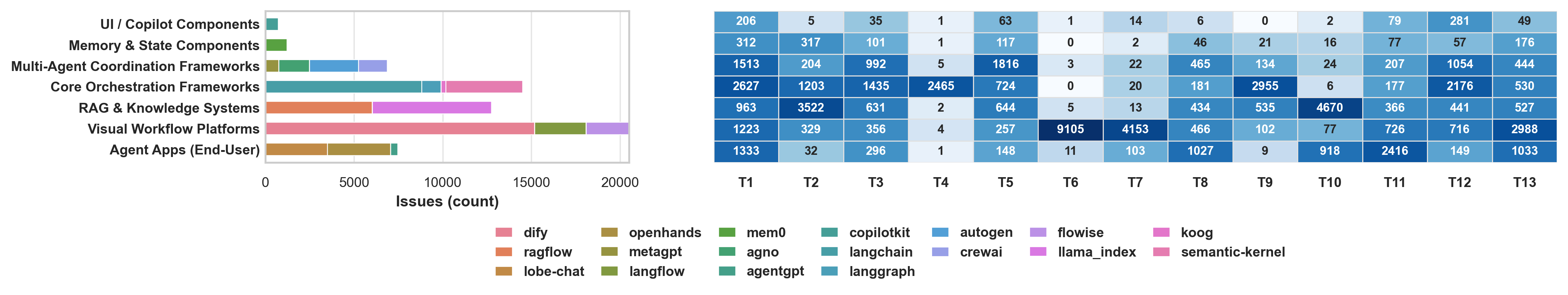}
    \caption{GitHub repo category topic counts}
    \label{fig:category_repo_topic_counts}
\end{figure*}

\begin{table}[t]
\caption{AI Agent technologies used on Stack Overflow.}
\label{tab:AIAgentTechnologies-compact}
\centering\scriptsize
\setlength{\tabcolsep}{3pt}
\renewcommand{\arraystretch}{0.92}
\begin{tabularx}{\columnwidth}{@{}>{\raggedright\arraybackslash}p{.40\columnwidth} >{\raggedright\arraybackslash}X@{}}
\toprule
\textbf{Category} & \textbf{Technologies (share \%)} \\
\midrule
\textbf{Orchestration frameworks} &
LangChain (70.6), LangGraph (4.7), CrewAI (1.5), AutoGen$^\dagger$ (2.3), Flowise (0.6), Semantic-Kernel (1.5) \\
\midrule
\textbf{Model APIs \& runtimes} &
OpenAI (24.2), Llama models/tooling (5.0), Ollama (3.8), Azure OpenAI (3.8),
Google GenAI (2.0), AutoGen$^\dagger$ (2.3), Groq (0.6), vLLM (0.6) \\
\midrule
\textbf{Retrieval \& Indexing} &
ChromaDB (9.6), LlamaIndex (4.1), FAISS (3.8), Pinecone (2.6), Weaviate (0.6) \\
\midrule
\textbf{Embedding Models} &
Hugging Face (5.8), Sentence-Transformers (1.5) \\
\midrule
\textbf{Evaluation} & RAGAS (0.6) \\
\midrule
\textbf{Search services} &
Azure Cognitive Search (2.0), Elasticsearch/OpenSearch (0.9) \\
\midrule
\textbf{Other} &
Streamlit (2.9), Neo4j (1.2), Anthropic (0.9), Azure AI Inference (0.3) \\
\bottomrule
\end{tabularx}

\vspace{0.25em}
\raggedright\footnotesize $^\dagger$Cross-listed: AutoGen appears under both Orchestration and Model APIs (used as an agentic framework and as a model/runtime interface).
\end{table}

\begin{table*}[t]
\caption{GitHub AI Agent Repositories.}
\label{tab:repo-metrics}
\centering\footnotesize
\setlength{\tabcolsep}{4pt}
\renewcommand{\arraystretch}{0.95}

\begin{tabularx}{\textwidth}{@{}
  >{\raggedright\arraybackslash}p{0.30\textwidth}
  >{\raggedright\arraybackslash}X
  r r r r
  >{\raggedright\arraybackslash}p{0.06\textwidth}
@{}}
\toprule
\textbf{Repository} & \textbf{Category} & \textbf{Stars} & \textbf{Forks} &
\textbf{\makecell{Open\\Issues}} & \textbf{\makecell{All\\Issues}} & \textbf{Lang} \\
\midrule
langgenius/dify                 & Visual Workflow Platforms      & 123{,}817 & 19{,}238 & 668  & 15{,}550 & TS \\
infiniflow/ragflow              & RAG \& Knowledge Systems       & 70{,}522  & 7{,}671  & 3{,}004 & 6{,}186  & Py \\
lobehub/lobe-chat               & Agent Apps (End-User)          & 69{,}535  & 14{,}288 & 1{,}158 & 4{,}984  & TS \\
OpenHands/OpenHands             & Agent Apps (End-User)          & 65{,}958  & 8{,}125  & 209  & 3{,}611  & Py \\
FoundationAgents/MetaGPT        & Multi-Agent Coordination       & 61{,}866  & 7{,}737  & 61   & 888      & Py \\
langflow-ai/langflow            & Visual Workflow Platforms      & 142{,}175 & 8{,}216  & 936  & 3{,}051  & Py \\
mem0ai/mem0                     & Memory \& State Components     & 44{,}730  & 4{,}864  & 573  & 1{,}295  & Py \\
agno-agi/agno                   & Multi-Agent Coordination       & 36{,}398  & 4{,}818  & 384  & 1{,}768  & Py \\
reworkd/AgentGPT                & Agent Apps (End-User)          & 35{,}405  & 9{,}465  & 217  & 469      & TS \\
CopilotKit/CopilotKit           & UI / Copilot Components        & 27{,}509  & 3{,}580  & 466  & 775      & TS \\
langchain-ai/langchain          & Core Orchestration             & 122{,}819 & 20{,}254 & 299  & 8{,}927  & Py \\
langchain-ai/langgraph          & Core Orchestration             & 22{,}626  & 3{,}977  & 276  & 1{,}107  & Py \\
microsoft/autogen               & Multi-Agent Coordination       & 52{,}913  & 8{,}040  & 532  & 2{,}910  & Py \\
crewAIInc/crewAI                & Multi-Agent Coordination       & 41{,}836  & 5{,}594  & 187  & 1{,}651  & Py \\
FlowiseAI/Flowise               & Visual Workflow Platforms      & 47{,}600  & 23{,}439 & 744  & 2{,}442  & TS \\
run-llama/llama\_index          & RAG \& Knowledge Systems       & 46{,}053  & 6{,}664  & 270  & 6{,}833  & Py \\
JetBrains/koog                  & Core Orchestration             & 3{,}570   & 283      & 186  & 279      & Kotlin \\
microsoft/semantic-kernel       & Core Orchestration             & 26{,}914  & 4{,}400  & 569  & 4{,}806  & C\# \\
\bottomrule
\end{tabularx}
\end{table*}

\begin{table}[t]
\caption{Language footprint: Stack Overflow usage vs.\ GitHub.}
\label{tab:lang-combined}
\centering\scriptsize
\setlength{\tabcolsep}{3pt}
\renewcommand{\arraystretch}{0.9}
\begin{tabular}{@{}l r r r@{}}
\toprule
\textbf{Language} & \textbf{SO share (\%)} & \textbf{Repo bytes} & \textbf{Repo share (\%)} \\
\midrule
Python            & 75.8 & 109{,}655{,}152 & 50.87 \\
TypeScript        & 5.5  & 57{,}855{,}653  & 26.84 \\
JavaScript        & 5.0  & 9{,}078{,}924   & 4.21  \\
C\#               & 0.3  & 17{,}487{,}007  & 8.11  \\
Java              & 3.2  & ---             & ---   \\
SQL               & 2.6  & ---             & ---   \\
Jupyter Notebook  & ---  & 10{,}380{,}564  & 4.82  \\
Kotlin            & ---  & 6{,}428{,}889   & 2.98  \\
\midrule
\textbf{Others / N\!A}$^{\dagger}$ & 12.2 & 4{,}684{,}636   & 2.17  \\
\bottomrule
\end{tabular}

\vspace{0.25em}
\raggedright\footnotesize $^{\dagger}$“Others / N\!A” aggregates Stack Overflow posts with unspecified language tags and GitHub languages outside the top six. Repo totals sum to 215{,}570{,}825 bytes.
\end{table}

\begin{table}[t]
\caption{Domain metrics by category in GitHub Issues (medians; d=days, h=hours).}
\centering
\setlength{\tabcolsep}{2.5pt}
\renewcommand{\arraystretch}{0.92}
\begin{tabular}{lrrrrr}
\toprule
Cat. & Open\% & Close d & Open d & Act h & Com \\
\midrule
UI/Copilot   & 52.0 & 14.1 & 166.5 & 477.3  & 3 \\
Memory/State & 28.7 & 13.7 & 177.2 & 279.3  & 2 \\
Multi-Agent  & 9.4  & 21.0 & 245.5 & 735.2  & 3 \\
Orch. Core   & 6.5  & 97.7 & 167.4 & 2350.5 & 2 \\
RAG/Knowl.   & 24.8 & 4.8  & 265.6 & 117.5  & 3 \\
Workflow Vis.& 6.9  & 1.7  & 172.3 & 116.6  & 2 \\
Agent Apps   & 14.7 & 4.8  & 258.3 & 1233.4 & 4 \\
\midrule
\textbf{Overall} & \textbf{12.5} & \textbf{9.2} & \textbf{218.8} & \textbf{484.1} & \textbf{2} \\
\bottomrule
\end{tabular}
\label{tab:cat_metrics_tiny}
\end{table}

\section{Discussion}
\paragraph{Evolution of AI Agent Challenges.}
Across both Stack Overflow and GitHub, the evolution of AI Agent challenges reflects a clear shift from early adoption hurdles to system-level complexity. In early 2023, Stack Overflow activity rises sharply around \emph{Installation \& Dependency Conflicts} (ST5), \emph{Document Embeddings \& Vector Stores} (ST2), and \emph{RAG Engineering} (ST6), indicating that developers initially struggled to assemble and operationalize emerging Agent stacks. This phase coincides with strong GitHub signals around \emph{Runtime Execution \& Operational Failures} (GT11) and \emph{Provider / Model Inconsistencies} (GT13), suggesting that early Agent frameworks were fragile and highly sensitive to environment configuration and provider behavior. From late 2023 into early 2024, several foundational challenges stabilize. Dependency-related Stack Overflow questions flatten, and GitHub issues related to runtime failures (GT11) peak and then decline, reflecting maturation of core orchestration frameworks and more reliable deployment practices. In contrast, \emph{Platforms \& Integrations} (GT1) remain persistently high throughout the entire period, highlighting that integrating Agents with external services, SDKs, and cloud environments remains a durable source of difficulty even as frameworks mature internally.
Beginning in mid-2024, the dominant challenges shift toward control, policy, and workflow semantics. GitHub issues related to \emph{Policy \& Template Enforcement Friction} (GT10) rise steadily, reflecting increasing emphasis on enforcing schemas, guardrails, and structured tool invocation in autonomous Agent loops. Most notably, \emph{Workflow, UI \& API Behavior} (GT6) becomes the single most prevalent topic through 2025, driven largely by Visual Workflow Platforms such as \texttt{Dify}, \texttt{LangFlow}, and \texttt{Flowise}. This surge indicates a transition toward UI-driven and low-code Agent development, where failures increasingly arise from workflow composition, API semantics, and user interaction rather than core model execution. Taken together, Stack Overflow captures early learning and configuration challenges, while GitHub reflects later-stage operational and ecosystem-level friction. Rather than diminishing over time, Agent difficulty shifts upward in the stack: as execution becomes more stable, challenges concentrate around orchestration, policy enforcement, and user-facing workflows, including application-level Agents used to compose and manage end-to-end workflows. This progression underscores that AI Agent complexity is rooted not only in models, but in autonomy, persistent state, and cross-system coordination at scale.

\paragraph{Technologies in Practice.}
Tables~\ref{tab:AIAgentTechnologies-compact} and~\ref{tab:repo-metrics} summarize the AI Agent technologies observed in our Stack Overflow and GitHub datasets. On GitHub, \texttt{Langflow}, \texttt{Dify}, and \texttt{LangChain} appear as the most prominent repositories in our sample, each exceeding 120K stars and associated with large volumes of issues related to orchestration, workflows, and integrations. Stack Overflow discussion is more concentrated, with \texttt{LangChain} and \texttt{LangGraph} accounting for the majority of Agent-related posts, followed by \texttt{LlamaIndex} and \texttt{AutoGen} (Table~\ref{tab:AIAgentTechnologies-compact}). While we examined additional Agent technologies identified on GitHub, many lacked stable Stack Overflow tags or sufficient post volume to support reliable, low-noise analysis. This reflects a Python-centric focus in Stack Overflow discussions, contrasted with greater language and platform diversity on GitHub (Table~\ref{tab:lang-combined}).

\begin{figure}
    \centering
    \includegraphics[width=1\linewidth]{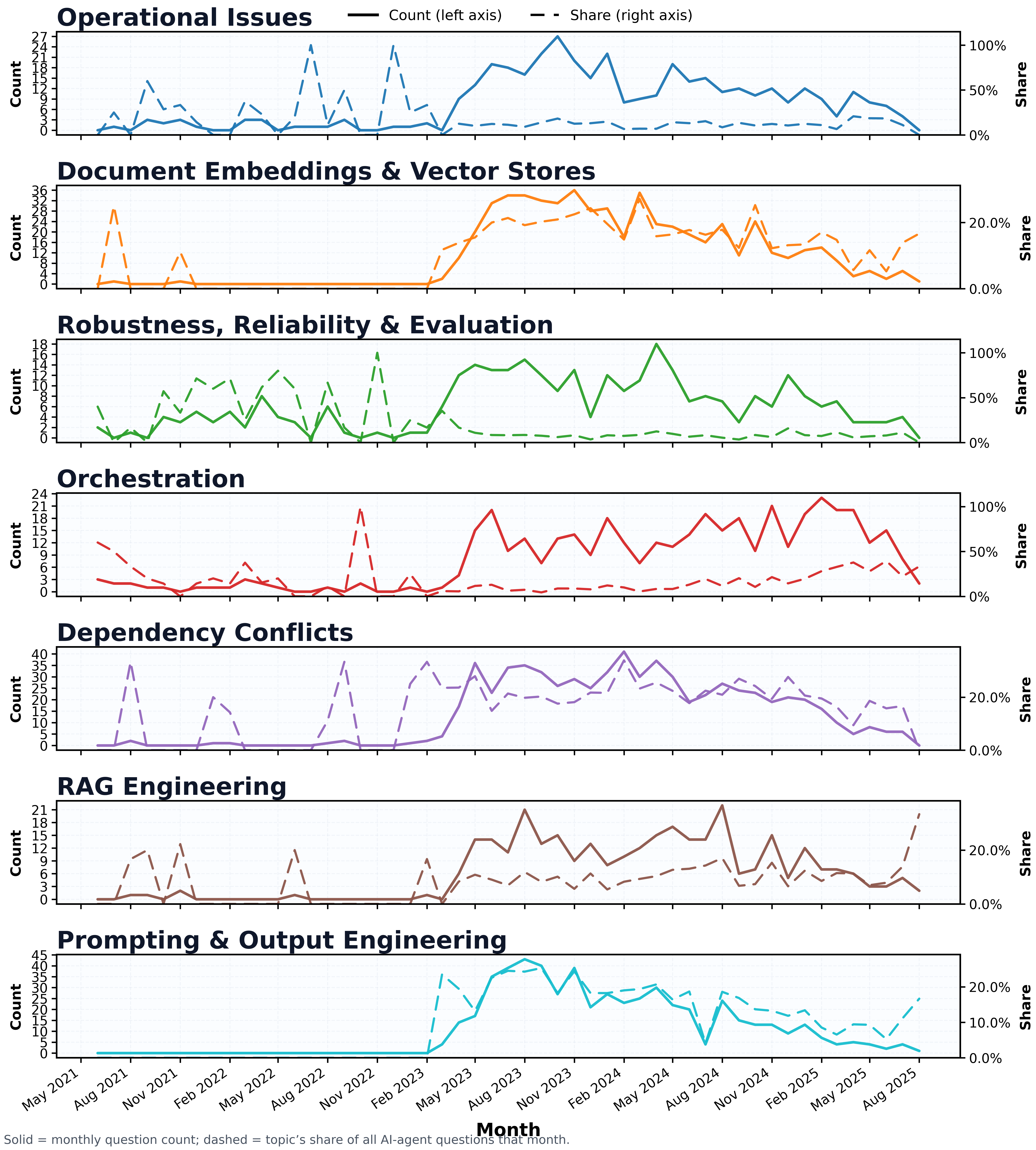}
    \caption{Stack Overflow AI Agent Topics Evolution.}
    \label{fig:topic_trends_small_multiples}
\end{figure}

\begin{figure}
    \centering
    \includegraphics[width=1\linewidth]{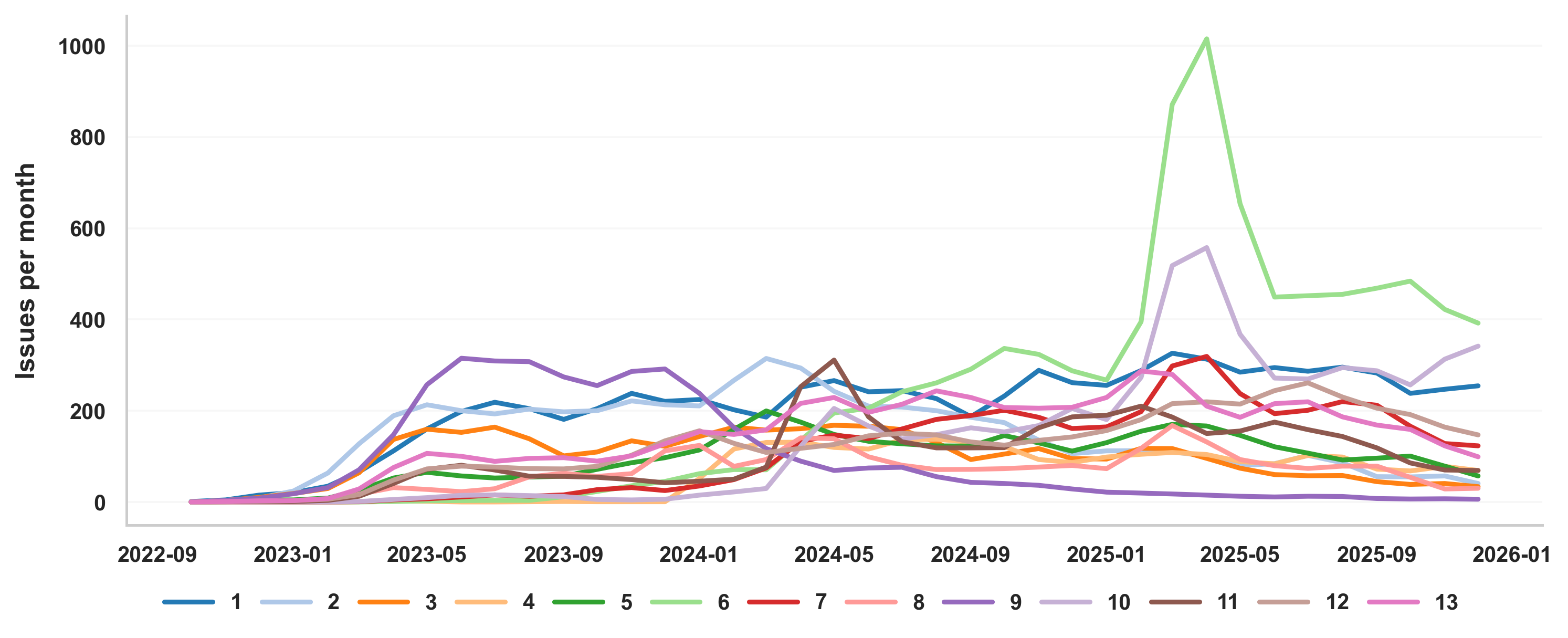}
    \caption{GitHub Issues topics overtime.}
    \label{fig:topics_overtime_paper_ready}
\end{figure}

\textbf{Comparing to Other Fields.}
As shown in Table~\ref{tab:popdiff_comparison_simple5} and prior studies \cite{ahmed2018concurrency,rosen2016mobile,yang2016security,bagherzadeh2019going,abdellatif2020challenges}, AI Agent topics are characterized by \emph{high difficulty}: they have the highest share of unanswered questions and the longest time-to-accept. While their popularity is comparable to established domains, fewer answers per question indicate slower community convergence. Overall, AI Agents combine strong engagement with above-average resolution difficulty.

\begin{table*}[t]
\caption{AI Agents vs.\ five prior Stack Overflow fields.}
\label{tab:popdiff_comparison_simple5}
\centering\footnotesize
\setlength{\tabcolsep}{4pt} 
\renewcommand{\arraystretch}{1.10}

\begin{tabularx}{\textwidth}{@{}%
>{\raggedright\arraybackslash}p{.18\textwidth}%
*{6}{>{\raggedleft\arraybackslash}p{.12\textwidth}}@{}}
\toprule
\textbf{Metric} & \textbf{AI Agents} & \textbf{Chatbot} & \textbf{Mobile} & \textbf{Security} & \textbf{Big Data} & \textbf{IoT} \\
\midrule
\textbf{(Popularity)}\\[-0.25em]
Avg.\ View & 2{,}240.1 & 512.4 & 2{,}300.0 & 2{,}461.1 & 1{,}560.4 & 1{,}320.3 \\
Avg.\ Score & 1.16 & 0.7 & 2.1 & 2.7 & 1.4 & 0.8 \\
Avg.\ AnswerCount & 0.93 & 1.0 & 1.5 & 1.6 & 1.1 & — \\
\addlinespace[0.35em]
\textbf{(Difficulty)}\\[-0.25em]
\% w/o Accepted Ans. & 82.6\% & 67.7\% & 52.0\% & 48.2\% & 60.3\% & 64.0\% \\
Median Hrs.\ to Accepted & 23.4 & 14.8 & 0.7 & 0.9 & 3.3 & 2.9 \\
\bottomrule
\end{tabularx}

\vspace{0.35em}
\raggedright\footnotesize
Notes: Prior-domain values reported from earlier SO studies \cite{ahmed2018concurrency, rosen2016mobile, yang2016security,  bagherzadeh2019going, abdellatif2020challenges}.
\end{table*}

\section{Related Work}

\textbf{AI Agents}. \ 
\cite{cemri2025multi} derive a data-driven taxonomy of multi-Agent LLM failures—\emph{specification}, \emph{inter-Agent misalignment}, \emph{task verification}—aligned to pre /execution/post stages via manual coding of $200{+}$ end-to-end traces. Different studies focus on distinct AI Agent challenges, including Security \cite{deng2025ai,zou2025security,de2025open}, Ethics \& governance \cite{gabriel2025we,kolt2025governing}, Foundations \& effectiveness of multi-Agent AI systems (MAS) \cite{tian2025outlook,du2021survey,du2025ai,sapkota2025ai}, and Design \& applications \cite{mo2025interactive,shetty2024building,cemri2025multi}. And also \cite{schneider2025generative} in a conceptual survey clarify how Agentic AI extends GenAI (reasoning, tools, memory, interaction), formalize agent specification, and catalog open challenges.

\textit{Security.} \cite{deng2025ai} survey the agent threat surface (intra-execution, agent–env, agent–agent, memory) and defenses (debate, RAG, constraints, post-correction, tool-use auditing). \cite{zou2025security} show near-universal, transferable prompt-injection/policy-violation attacks in large-scale red teaming, exposing defense gaps. \cite{de2025open} argue for \emph{multi-agent security} as a distinct field, outlining open problems (collusion, attribution, cascade dynamics) and a research agenda. \textit{Ethics \& governance.} \cite{gabriel2025we} call for new ethics for autonomous agents—safety, alignment, and social-coordination risks as autonomy grows. \cite{kolt2025governing} frame governance as a principal–agent problem and advocate transparency, incentive design, monitoring, and accountability across developers/deployers/users. \textit{Foundations \& effectiveness of MAS.} \cite{tian2025outlook} formalize multi-agent AI, showing gains from task reallocation and diverse agents, but error amplification under overlap/misalignment; emphasize feedback that reshapes topology. \cite{du2021survey} survey multi-agent deep reinforcement learning, stressing non-stationarity, partial observability, scalability, and credit assignment. \cite{du2025ai} review agent–agent communication, distilling five pillars (scalability, security, real-time, performance, manageability) and standardization paths. \textit{Design \& applications.} \cite{mo2025interactive} study an interactive refactoring agent, reporting decision strategies and collaboration principles. \cite{shetty2024building} sketch requirements and a prototype (AIOpsLab) for a standardized framework to build/test/compare cloud-ops agents.

\section{Implications and Recommendations}

Our findings suggest that the primary difficulties in AI Agent development do not stem from model capability alone, but from the interaction between autonomous control flow, persistent state, tool invocation, and user-facing workflows. Based on the observed challenge distributions across Stack Overflow, GitHub, ecosystem domains, and time, we derive the following implications for Agent designers, framework builders, and researchers.

\textbf{(1) Treat orchestration as a first-class architectural concern.}
Across both Stack Overflow (ST4) and GitHub (GT12), orchestration-related challenges are consistently among the most difficult, despite relatively lower visibility on Q\&A platforms. This aligns strongly with \emph{Planning} and \emph{Multi-Agent} design patterns, where Agents explicitly decompose goals into sub-tasks, route control between steps or sub-Agents, and coordinate execution over time rather than issuing a single model call. Failures in these patterns typically arise from task decomposition, routing, and execution order rather than individual model outputs. Frameworks should therefore expose orchestration primitives explicitly—state transitions, retries, and termination conditions—rather than embedding them implicitly inside chains or graphs. Clear execution semantics and debuggable control flow are essential for scaling beyond single-step Agents.

\textbf{(2) Stabilize interaction contracts between models, tools, and memory.}
A large fraction of unresolved issues concentrate around structured inputs and outputs (ST7, GT3, GT10), reflecting fragility in prompt schemas, tool-calling formats, and validation rules. These challenges directly affect \emph{ReAct} and \emph{Tool Use} patterns, in which an Agent alternates between reasoning steps and external actions (e.g., API calls, database queries), and where even minor contract violations can halt entire execution loops. The rise of Model Context Protocol (MCP)–style abstractions is a promising direction: explicit, versioned contracts for context, tools, and memory can reduce silent failures and make Agent behavior more predictable across providers and environments.

\textbf{(3) Design memory and retrieval as evolving systems, not static components.}
Retrieval- and memory-related topics (ST2, ST6, GT2, GT13) are consistently among the hardest to resolve and show long-lived issue activity. This reflects the mismatch between static RAG pipelines and the needs of long-running, stateful Agents. Patterns such as \emph{Reflection} and \emph{Planning} assume that Agents can reliably store intermediate results, revisit prior decisions, and update internal state over time—assumptions that current tooling often fails to guarantee. We therefore recommend treating Agent memory as a lifecycle-managed subsystem, with explicit policies for updating, pruning, and validating state, rather than as a passive vector store queried once per task.

\textbf{(4) Expect UI- and workflow-centric Agents to dominate future development effort.}
Over time, GitHub activity increasingly concentrates in \emph{Workflow, UI \& API Behavior} (GT6), particularly within visual workflow platforms such as Dify, Langflow, and Flowise. These systems operationalize Agentic design patterns by allowing users to compose planning, tool use, and multi-step execution through graphical or declarative workflows, often to build other Agents or application-level pipelines. While many UI-related issues resolve quickly, their sheer volume indicates that the next phase of Agent adoption is driven by end-user composition rather than low-level model access. This suggests that Agent frameworks must prioritize explainability, safe defaults, and recoverability at the workflow level.

\textbf{(5) Application-level Agents amplify upstream design decisions.}
Agent applications aggregate failures from across the stack, including runtime execution (GT11), platform integrations (GT1), provider inconsistencies (GT13), and unhandled exceptions (GT8). In design-pattern terms, application-level Agents frequently combine \emph{Tool Use}, \emph{Planning}, and \emph{Multi-Agent} behaviors into a single system, making them especially sensitive to upstream instability. Our results indicate that robustness at the application layer depends less on adding local safeguards and more on improving guarantees in orchestration, interaction contracts, and memory management below.

\textbf{Implications for Practitioners.}
For practitioners building and deploying AI Agents, our findings emphasize that most failures originate at system boundaries rather than within individual model calls. Issues related to orchestration, runtime execution, and workflow behavior dominate GitHub activity, especially in application-level Agents and visual workflow platforms. Practitioners should therefore prioritize explicit control flow, strong observability, and defensive execution strategies over increasing Agent autonomy. In practice, this means treating planning loops, retries, and tool invocation as auditable system components, and investing early in workflow-level safeguards when deploying user-facing or application-based Agents.

\textbf{Implications for Educators.}
For educators, the results reveal a gap between what learners most frequently encounter and what they find hardest to master. While installation, prompting, and basic RAG pipelines attract the most attention, the most difficult challenges involve orchestration, persistent memory, and long-running execution. Educational material should therefore move beyond linear prompt–response examples and instead teach Agent behavior over time, including failure modes, execution traces, and debugging practices across Agentic design patterns such as ReAct, Planning, and Multi-Agent coordination.

\textbf{Implications for Researchers.}
For researchers, the findings suggest that progress in AI Agents depends less on improving standalone model capabilities and more on formalizing system-level abstractions. Persistent difficulty in orchestration, retrieval, and interaction contracts indicates a need for research on execution semantics, memory lifecycle management, and standardized interfaces between models and tools. Emerging efforts such as Model Context Protocols point toward promising directions, but our data shows that these abstractions must be evaluated under long-running, user-facing conditions. Future research should therefore prioritize reproducible Agent failures, coordination mechanisms, and cross-component guarantees rather than isolated reasoning benchmarks.

\textbf{Overall Implication.}
Taken together, these findings indicate that AI Agent development is transitioning from a model-centric phase to a systems-engineering phase. The dominant challenges now arise from autonomy, persistent state, and cross-component coordination rather than prompt quality alone. Agentic design patterns remain useful abstractions, but their practical success increasingly depends on explicit orchestration semantics, stable interaction protocols, and tooling that supports long-running, user-facing workflows.

\section{Threats To Validity}

    \textbf{Selection of tags.} Relying on tags to collect Stack Overflow questions and answers is a validity threat, as tagging can be incomplete, and some relevant posts may be missed. To reduce this risk, we followed well-established techniques \cite{ahmed2018concurrency,rosen2016mobile,yang2016security, haque2020challenges}. However, the creation of our tag set could introduce bias, as the selected tags may not cover all AI Agent–related questions, and the final tag set could be influenced by individual experiences. \textcolor{black}{To mitigate this, we engaged the first two researchers in the validation process to enhance the reliability and validity of the final tag selection.}

\textbf{Manual labeling}. We manually reviewed a stratified, confidence-aware sample (unlike purely random sampling in much prior work), weighting by each topic’s size and the model’s per-record confidence, and assigned topics labels following established protocols \cite{sayyadnejad2024exploring,uddin2021empirical,mahmood2023empirical,chen2020comprehensive}; while some bias is possible, \textcolor{black}{two authors performed analyzing and reconciliation, and a third author independently validated the final labels.}
\vspace{-5pt}

\section{Conclusion}

This study presents a large-scale study on technical challenges developers face when constructing AI Agents, based on large-scale data from Stack Overflow and GitHub. The findings indicate that Agent development is primarily a systems-engineering challenge rather than solely a limitation of model capability. Analysis of Stack Overflow data identifies \textbf{seven recurring Agent challenge} topics, further divided into \textbf{28 fine-grained subtopics}, encompassing installation, retrieval, orchestration, interaction contracts, and runtime reliability. A complementary examination of GitHub issues reveals\textbf{ 13 Agent-related challenges}, illustrating how these problems arise during framework development, integration, and maintenance. Across both platforms, configuration and prompting issues receive the most attention and are typically resolved rapidly, while challenges related to retrieval pipelines, orchestration semantics, and long-running execution remain the most difficult to address. We noticed the most challenging Agent issues are not the most frequently discussed and tend to have higher rates of unresolved questions and longer resolution times. Domain-level analysis further revealed infrastructure-oriented domains accumulate fewer but longer-running issues, whereas UI- and workflow-centric platforms generate a higher volume of faster-resolving reports. Application-level Agents amplify failures throughout the stack, demonstrating that upstream design decisions in orchestration, memory, and interaction contracts propagate to user-facing systems. Progress in AI Agents relies less on improving individual models and more on enhancing orchestration mechanisms, stabilizing interaction contracts, and managing persistent state across complex, tool-mediated workflows.



\bibliography{software} 

\begin{thebibliography}{52}
\expandafter\ifx\csname natexlab\endcsname\relax\def\natexlab#1{#1}\fi
\providecommand{\url}[1]{\texttt{#1}}
\providecommand{\href}[2]{#2}
\providecommand{\path}[1]{#1}
\providecommand{\DOIprefix}{doi:}
\providecommand{\ArXivprefix}{arXiv:}
\providecommand{\URLprefix}{URL: }
\providecommand{\Pubmedprefix}{pmid:}
\providecommand{\doi}[1]{\href{http://dx.doi.org/#1}{\path{#1}}}
\providecommand{\Pubmed}[1]{\href{pmid:#1}{\path{#1}}}
\providecommand{\bibinfo}[2]{#2}
\ifx\xfnm\relax \def\xfnm[#1]{\unskip,\space#1}\fi
\bibitem[{Abdellatif et~al.(2020)Abdellatif, Costa, Badran, Abdalkareem and Shihab}]{abdellatif2020challenges}
\bibinfo{author}{Abdellatif, A.}, \bibinfo{author}{Costa, D.}, \bibinfo{author}{Badran, K.}, \bibinfo{author}{Abdalkareem, R.}, \bibinfo{author}{Shihab, E.}, \bibinfo{year}{2020}.
\newblock \bibinfo{title}{Challenges in chatbot development: A study of stack overflow posts}, in: \bibinfo{booktitle}{Proceedings of the 17th international conference on mining software repositories}, pp. \bibinfo{pages}{174--185}.
\bibitem[{Ahmed et~al.(2025)Ahmed, Opu, Roy, Suhi and Chowdhury}]{ahmed2025exploring}
\bibinfo{author}{Ahmed, M.}, \bibinfo{author}{Opu, M.N.I.}, \bibinfo{author}{Roy, C.}, \bibinfo{author}{Suhi, S.I.}, \bibinfo{author}{Chowdhury, S.}, \bibinfo{year}{2025}.
\newblock \bibinfo{title}{Exploring challenges in test mocking: Developer questions and insights from stackoverflow}.
\newblock \bibinfo{journal}{arXiv preprint arXiv:2505.08300} .
\bibitem[{Ahmed and Bagherzadeh(2018)}]{ahmed2018concurrency}
\bibinfo{author}{Ahmed, S.}, \bibinfo{author}{Bagherzadeh, M.}, \bibinfo{year}{2018}.
\newblock \bibinfo{title}{What do concurrency developers ask about? a large-scale study using stack overflow}, in: \bibinfo{booktitle}{Proceedings of the 12th ACM/IEEE international symposium on empirical software engineering and measurement}, pp. \bibinfo{pages}{1--10}.
\bibitem[{Akbarpour et~al.(2025)Akbarpour, Mirza, Raoofian, Fard and Rodr{\'\i}guez-P{\'e}rez}]{akbarpour2025unveiling}
\bibinfo{author}{Akbarpour, N.}, \bibinfo{author}{Mirza, A.S.}, \bibinfo{author}{Raoofian, E.}, \bibinfo{author}{Fard, F.}, \bibinfo{author}{Rodr{\'\i}guez-P{\'e}rez, G.}, \bibinfo{year}{2025}.
\newblock \bibinfo{title}{Unveiling ruby: Insights from stack overflow and developer survey}.
\newblock \bibinfo{journal}{arXiv preprint arXiv:2503.19238} .
\bibitem[{Alamin et~al.(2023)Alamin, Uddin, Malakar, Afroz, Haider and Iqbal}]{alamin2023developer}
\bibinfo{author}{Alamin, M.A.A.}, \bibinfo{author}{Uddin, G.}, \bibinfo{author}{Malakar, S.}, \bibinfo{author}{Afroz, S.}, \bibinfo{author}{Haider, T.}, \bibinfo{author}{Iqbal, A.}, \bibinfo{year}{2023}.
\newblock \bibinfo{title}{Developer discussion topics on the adoption and barriers of low code software development platforms}.
\newblock \bibinfo{journal}{Empirical software engineering} \bibinfo{volume}{28}, \bibinfo{pages}{4}.
\bibitem[{Allamanis and Sutton(2013)}]{allamanis2013and}
\bibinfo{author}{Allamanis, M.}, \bibinfo{author}{Sutton, C.}, \bibinfo{year}{2013}.
\newblock \bibinfo{title}{Why, when, and what: analyzing stack overflow questions by topic, type, and code}, in: \bibinfo{booktitle}{2013 10th Working conference on mining software repositories (MSR)}, \bibinfo{organization}{IEEE}. pp. \bibinfo{pages}{53--56}.
\bibitem[{Asgari et~al.(2024)Asgari, Guerriero, Pietrantuono and Russo}]{asgari2024testing}
\bibinfo{author}{Asgari, A.}, \bibinfo{author}{Guerriero, A.}, \bibinfo{author}{Pietrantuono, R.}, \bibinfo{author}{Russo, S.}, \bibinfo{year}{2024}.
\newblock \bibinfo{title}{From testing to evaluation of nlp and llm systems: An analysis of researchers and practitioners perspectives through systematic literature review and developers’ community platforms mining} .
\bibitem[{Asgari et~al.()Asgari, Guerriero, Pietrantuono, Russo et~al.}]{asgari2025adaptive}
\bibinfo{author}{Asgari, A.}, \bibinfo{author}{Guerriero, A.}, \bibinfo{author}{Pietrantuono, R.}, \bibinfo{author}{Russo, S.}, et~al., .
\newblock \bibinfo{title}{Adaptive probabilistic operational testing for large language models evaluation}.
\bibitem[{Bagherzadeh and Khatchadourian(2019)}]{bagherzadeh2019going}
\bibinfo{author}{Bagherzadeh, M.}, \bibinfo{author}{Khatchadourian, R.}, \bibinfo{year}{2019}.
\newblock \bibinfo{title}{Going big: a large-scale study on what big data developers ask}, in: \bibinfo{booktitle}{Proceedings of the 2019 27th ACM joint meeting on european software engineering conference and symposium on the foundations of software engineering}, pp. \bibinfo{pages}{432--442}.
\bibitem[{Barua et~al.(2014)Barua, Thomas and Hassan}]{barua2014developers}
\bibinfo{author}{Barua, A.}, \bibinfo{author}{Thomas, S.W.}, \bibinfo{author}{Hassan, A.E.}, \bibinfo{year}{2014}.
\newblock \bibinfo{title}{What are developers talking about? an analysis of topics and trends in stack overflow}.
\newblock \bibinfo{journal}{Empirical software engineering} \bibinfo{volume}{19}, \bibinfo{pages}{619--654}.
\bibitem[{Bi et~al.(2021)Bi, Liang, Tang and Xia}]{bi2021mining}
\bibinfo{author}{Bi, T.}, \bibinfo{author}{Liang, P.}, \bibinfo{author}{Tang, A.}, \bibinfo{author}{Xia, X.}, \bibinfo{year}{2021}.
\newblock \bibinfo{title}{Mining architecture tactics and quality attributes knowledge in stack overflow}.
\newblock \bibinfo{journal}{Journal of Systems and Software} \bibinfo{volume}{180}, \bibinfo{pages}{111005}.
\bibitem[{Braiek et~al.(2018)Braiek, Khomh and Adams}]{braiek2018open}
\bibinfo{author}{Braiek, H.B.}, \bibinfo{author}{Khomh, F.}, \bibinfo{author}{Adams, B.}, \bibinfo{year}{2018}.
\newblock \bibinfo{title}{The open-closed principle of modern machine learning frameworks}, in: \bibinfo{booktitle}{Proceedings of the 15th international conference on mining software repositories}, pp. \bibinfo{pages}{353--363}.
\bibitem[{Cemri et~al.(2025)Cemri, Pan, Yang, Agrawal, Chopra, Tiwari, Keutzer, Parameswaran, Klein, Ramchandran et~al.}]{cemri2025multi}
\bibinfo{author}{Cemri, M.}, \bibinfo{author}{Pan, M.Z.}, \bibinfo{author}{Yang, S.}, \bibinfo{author}{Agrawal, L.A.}, \bibinfo{author}{Chopra, B.}, \bibinfo{author}{Tiwari, R.}, \bibinfo{author}{Keutzer, K.}, \bibinfo{author}{Parameswaran, A.}, \bibinfo{author}{Klein, D.}, \bibinfo{author}{Ramchandran, K.}, et~al., \bibinfo{year}{2025}.
\newblock \bibinfo{title}{Why do multi-agent llm systems fail?}
\newblock \bibinfo{journal}{arXiv preprint arXiv:2503.13657} .
\bibitem[{Chen et~al.(2020)Chen, Cao, Liu, Wang, Xie and Liu}]{chen2020comprehensive}
\bibinfo{author}{Chen, Z.}, \bibinfo{author}{Cao, Y.}, \bibinfo{author}{Liu, Y.}, \bibinfo{author}{Wang, H.}, \bibinfo{author}{Xie, T.}, \bibinfo{author}{Liu, X.}, \bibinfo{year}{2020}.
\newblock \bibinfo{title}{A comprehensive study on challenges in deploying deep learning based software}, in: \bibinfo{booktitle}{Proceedings of the 28th ACM joint meeting on European software engineering conference and symposium on the foundations of software engineering}, pp. \bibinfo{pages}{750--762}.
\bibitem[{Cochran(1977)}]{cochran1977sampling}
\bibinfo{author}{Cochran, W.G.}, \bibinfo{year}{1977}.
\newblock \bibinfo{title}{Sampling techniques}.
\newblock \bibinfo{publisher}{john wiley \& sons}.
\bibitem[{Deng et~al.(2025)Deng, Guo, Han, Ma, Xiong, Wen and Xiang}]{deng2025ai}
\bibinfo{author}{Deng, Z.}, \bibinfo{author}{Guo, Y.}, \bibinfo{author}{Han, C.}, \bibinfo{author}{Ma, W.}, \bibinfo{author}{Xiong, J.}, \bibinfo{author}{Wen, S.}, \bibinfo{author}{Xiang, Y.}, \bibinfo{year}{2025}.
\newblock \bibinfo{title}{Ai agents under threat: A survey of key security challenges and future pathways}.
\newblock \bibinfo{journal}{ACM Computing Surveys} \bibinfo{volume}{57}, \bibinfo{pages}{1--36}.
\bibitem[{Du et~al.(2025)Du, Wang, Chao, Xie and Cui}]{du2025ai}
\bibinfo{author}{Du, C.}, \bibinfo{author}{Wang, C.}, \bibinfo{author}{Chao, Y.}, \bibinfo{author}{Xie, X.}, \bibinfo{author}{Cui, Y.}, \bibinfo{year}{2025}.
\newblock \bibinfo{title}{Ai agent communication from internet architecture perspective: Challenges and opportunities}.
\newblock \bibinfo{journal}{arXiv preprint arXiv:2509.02317} .
\bibitem[{Du and Ding(2021)}]{du2021survey}
\bibinfo{author}{Du, W.}, \bibinfo{author}{Ding, S.}, \bibinfo{year}{2021}.
\newblock \bibinfo{title}{A survey on multi-agent deep reinforcement learning: from the perspective of challenges and applications}.
\newblock \bibinfo{journal}{Artificial Intelligence Review} \bibinfo{volume}{54}, \bibinfo{pages}{3215--3238}.
\bibitem[{Elshan et~al.(2024)Elshan, Bruhin, Schmidt et~al.}]{elshan2024unveiling}
\bibinfo{author}{Elshan, E.}, \bibinfo{author}{Bruhin, O.}, \bibinfo{author}{Schmidt, N.}, et~al., \bibinfo{year}{2024}.
\newblock \bibinfo{title}{Unveiling challenges and opportunities in low code development platforms: A stackoverflow analysis}, in: \bibinfo{booktitle}{Proceedings of the 57th Hawaii International Conference on System Sciences (HICSS)}.
\bibitem[{Fincher and Tenenberg(2005)}]{fincher2005making}
\bibinfo{author}{Fincher, S.}, \bibinfo{author}{Tenenberg, J.}, \bibinfo{year}{2005}.
\newblock \bibinfo{title}{Making sense of card sorting data}.
\newblock \bibinfo{journal}{Expert Systems} \bibinfo{volume}{22}, \bibinfo{pages}{89--93}.
\bibitem[{Gabriel et~al.(2025)Gabriel, Keeling, Manzini and Evans}]{gabriel2025we}
\bibinfo{author}{Gabriel, I.}, \bibinfo{author}{Keeling, G.}, \bibinfo{author}{Manzini, A.}, \bibinfo{author}{Evans, J.}, \bibinfo{year}{2025}.
\newblock \bibinfo{title}{We need a new ethics for a world of ai agents}.
\newblock \bibinfo{journal}{arXiv preprint arXiv:2509.10289} .
\bibitem[{Han et~al.(2020)Han, Shihab, Wan, Deng and Xia}]{han2020programmers}
\bibinfo{author}{Han, J.}, \bibinfo{author}{Shihab, E.}, \bibinfo{author}{Wan, Z.}, \bibinfo{author}{Deng, S.}, \bibinfo{author}{Xia, X.}, \bibinfo{year}{2020}.
\newblock \bibinfo{title}{What do programmers discuss about deep learning frameworks}.
\newblock \bibinfo{journal}{Empirical Software Engineering} \bibinfo{volume}{25}, \bibinfo{pages}{2694--2747}.
\bibitem[{Haque et~al.(2020)Haque, Iwaya and Babar}]{haque2020challenges}
\bibinfo{author}{Haque, M.U.}, \bibinfo{author}{Iwaya, L.H.}, \bibinfo{author}{Babar, M.A.}, \bibinfo{year}{2020}.
\newblock \bibinfo{title}{Challenges in docker development: A large-scale study using stack overflow}, in: \bibinfo{booktitle}{Proceedings of the 14th ACM/IEEE international symposium on empirical software engineering and measurement (ESEM)}, pp. \bibinfo{pages}{1--11}.
\bibitem[{Hindle et~al.(2016)Hindle, Alipour and Stroulia}]{hindle2016contextual}
\bibinfo{author}{Hindle, A.}, \bibinfo{author}{Alipour, A.}, \bibinfo{author}{Stroulia, E.}, \bibinfo{year}{2016}.
\newblock \bibinfo{title}{A contextual approach towards more accurate duplicate bug report detection and ranking}.
\newblock \bibinfo{journal}{Empirical Software Engineering} \bibinfo{volume}{21}, \bibinfo{pages}{368--410}.
\bibitem[{Honnibal et~al.(2020)Honnibal, Montani, Van~Landeghem, Boyd et~al.}]{honnibal2020spacy}
\bibinfo{author}{Honnibal, M.}, \bibinfo{author}{Montani, I.}, \bibinfo{author}{Van~Landeghem, S.}, \bibinfo{author}{Boyd, A.}, et~al., \bibinfo{year}{2020}.
\newblock \bibinfo{title}{spacy: Industrial-strength natural language processing in python} .
\bibitem[{Kolt(2025)}]{kolt2025governing}
\bibinfo{author}{Kolt, N.}, \bibinfo{year}{2025}.
\newblock \bibinfo{title}{Governing ai agents}.
\newblock \bibinfo{journal}{arXiv preprint arXiv:2501.07913} .
\bibitem[{Kubiak and Kawalec(2022)}]{kubiak2022prior}
\bibinfo{author}{Kubiak, A.P.}, \bibinfo{author}{Kawalec, P.}, \bibinfo{year}{2022}.
\newblock \bibinfo{title}{Prior information in frequentist research designs: The case of neyman’s sampling theory}.
\newblock \bibinfo{journal}{Journal for General Philosophy of Science} \bibinfo{volume}{53}, \bibinfo{pages}{381--402}.
\bibitem[{Li et~al.(2021)Li, Khomh, Openja et~al.}]{li2021understanding}
\bibinfo{author}{Li, H.}, \bibinfo{author}{Khomh, F.}, \bibinfo{author}{Openja, M.}, et~al., \bibinfo{year}{2021}.
\newblock \bibinfo{title}{Understanding quantum software engineering challenges an empirical study on stack exchange forums and github issues}, in: \bibinfo{booktitle}{2021 IEEE International Conference on Software Maintenance and Evolution (ICSME)}, \bibinfo{organization}{IEEE}. pp. \bibinfo{pages}{343--354}.
\bibitem[{Loper and Bird(2002)}]{loper2002nltk}
\bibinfo{author}{Loper, E.}, \bibinfo{author}{Bird, S.}, \bibinfo{year}{2002}.
\newblock \bibinfo{title}{Nltk: The natural language toolkit}.
\newblock \bibinfo{journal}{arXiv preprint cs/0205028} .
\bibitem[{Mahmood et~al.(2023)Mahmood, Rasool, Sabir and Athar}]{mahmood2023empirical}
\bibinfo{author}{Mahmood, K.}, \bibinfo{author}{Rasool, G.}, \bibinfo{author}{Sabir, F.}, \bibinfo{author}{Athar, A.}, \bibinfo{year}{2023}.
\newblock \bibinfo{title}{An empirical study of web services topics in web developer discussions on stack overflow}.
\newblock \bibinfo{journal}{IEEE Access} \bibinfo{volume}{11}, \bibinfo{pages}{9627--9655}.
\bibitem[{McCallum(2002)}]{mccallum2002mallet}
\bibinfo{author}{McCallum, A.K.}, \bibinfo{year}{2002}.
\newblock \bibinfo{title}{Mallet: A machine learning for languagetoolkit}.
\newblock \bibinfo{journal}{http://mallet. cs. umass. edu} .
\bibitem[{McHugh(2012)}]{mchugh2012interrater}
\bibinfo{author}{McHugh, M.L.}, \bibinfo{year}{2012}.
\newblock \bibinfo{title}{Interrater reliability: the kappa statistic}.
\newblock \bibinfo{journal}{Biochemia medica} \bibinfo{volume}{22}, \bibinfo{pages}{276--282}.
\bibitem[{Mo et~al.(2025)Mo, Jiang and Zheng}]{mo2025interactive}
\bibinfo{author}{Mo, T.}, \bibinfo{author}{Jiang, Z.}, \bibinfo{author}{Zheng, Q.}, \bibinfo{year}{2025}.
\newblock \bibinfo{title}{Interactive ai agent for code refactoring assistance: A study on decision-making strategies and human-agent collaboration effectiveness}.
\newblock \bibinfo{journal}{Academia Nexus Journal} \bibinfo{volume}{4}.
\bibitem[{Neyman(1938)}]{neyman1938contribution}
\bibinfo{author}{Neyman, J.}, \bibinfo{year}{1938}.
\newblock \bibinfo{title}{Contribution to the theory of sampling human populations}.
\newblock \bibinfo{journal}{Journal of the American Statistical Association} \bibinfo{volume}{33}, \bibinfo{pages}{101--116}.
\bibitem[{Panichella et~al.(2013)Panichella, Dit, Oliveto, Di~Penta, Poshynanyk and De~Lucia}]{panichella2013effectively}
\bibinfo{author}{Panichella, A.}, \bibinfo{author}{Dit, B.}, \bibinfo{author}{Oliveto, R.}, \bibinfo{author}{Di~Penta, M.}, \bibinfo{author}{Poshynanyk, D.}, \bibinfo{author}{De~Lucia, A.}, \bibinfo{year}{2013}.
\newblock \bibinfo{title}{How to effectively use topic models for software engineering tasks? an approach based on genetic algorithms}, in: \bibinfo{booktitle}{2013 35th International Conference on Software Engineering (ICSE)}, pp. \bibinfo{pages}{522--531}.
\newblock \DOIprefix\doi{10.1109/ICSE.2013.6606598}.
\bibitem[{{\v{R}}eh{\r{u}}{\v{r}}ek and Sojka(2010)}]{vrehuuvrek2010software}
\bibinfo{author}{{\v{R}}eh{\r{u}}{\v{r}}ek, R.}, \bibinfo{author}{Sojka, P.}, \bibinfo{year}{2010}.
\newblock \bibinfo{title}{Software framework for topic modelling with large corpora} .
\bibitem[{Renthlei and Lallawmkima()}]{renthleichoosing}
\bibinfo{author}{Renthlei, Z.}, \bibinfo{author}{Lallawmkima, C.}, .
\newblock \bibinfo{title}{Choosing the right sample size in social science studies: A methodological review}.
\newblock \bibinfo{journal}{Mizoram Educational Journal} , \bibinfo{pages}{1}.
\bibitem[{R{\"o}der et~al.(2015)R{\"o}der, Both and Hinneburg}]{roder2015exploring}
\bibinfo{author}{R{\"o}der, M.}, \bibinfo{author}{Both, A.}, \bibinfo{author}{Hinneburg, A.}, \bibinfo{year}{2015}.
\newblock \bibinfo{title}{Exploring the space of topic coherence measures}, in: \bibinfo{booktitle}{Proceedings of the eighth ACM international conference on Web search and data mining}, pp. \bibinfo{pages}{399--408}.
\bibitem[{Rosen and Shihab(2016)}]{rosen2016mobile}
\bibinfo{author}{Rosen, C.}, \bibinfo{author}{Shihab, E.}, \bibinfo{year}{2016}.
\newblock \bibinfo{title}{What are mobile developers asking about? a large scale study using stack overflow}.
\newblock \bibinfo{journal}{Empirical Software Engineering} \bibinfo{volume}{21}, \bibinfo{pages}{1192--1223}.
\bibitem[{Sapkota et~al.(2025)Sapkota, Roumeliotis and Karkee}]{sapkota2025ai}
\bibinfo{author}{Sapkota, R.}, \bibinfo{author}{Roumeliotis, K.I.}, \bibinfo{author}{Karkee, M.}, \bibinfo{year}{2025}.
\newblock \bibinfo{title}{Ai agents vs. agentic ai: A conceptual taxonomy, applications and challenges}.
\newblock \bibinfo{journal}{arXiv preprint arXiv:2505.10468} .
\bibitem[{Sayyadnejad et~al.(2025)Sayyadnejad, Asgari, Sami and Tahayori}]{sayyadnejad2024exploring}
\bibinfo{author}{Sayyadnejad, M.M.}, \bibinfo{author}{Asgari, A.}, \bibinfo{author}{Sami, A.}, \bibinfo{author}{Tahayori, H.}, \bibinfo{year}{2025}.
\newblock \bibinfo{title}{Exploring the black box: analysing explainable ai challenges and best practices through stack exchange discussions}.
\newblock \bibinfo{journal}{Empirical Software Engineering} \bibinfo{volume}{30}, \bibinfo{pages}{176}.
\bibitem[{Schneider(2025)}]{schneider2025generative}
\bibinfo{author}{Schneider, J.}, \bibinfo{year}{2025}.
\newblock \bibinfo{title}{Generative to agentic ai: Survey, conceptualization, and challenges}.
\newblock \bibinfo{journal}{arXiv preprint arXiv:2504.18875} .
\bibitem[{Sengupta and Haythornthwaite(2020)}]{sengupta2020learning}
\bibinfo{author}{Sengupta, S.}, \bibinfo{author}{Haythornthwaite, C.}, \bibinfo{year}{2020}.
\newblock \bibinfo{title}{Learning with comments: An analysis of comments and community on stack overflow} .
\bibitem[{Shetty et~al.(2024)Shetty, Chen, Somashekar, Ma, Simmhan, Zhang, Mace, Vandevoorde, Las-Casas, Gupta et~al.}]{shetty2024building}
\bibinfo{author}{Shetty, M.}, \bibinfo{author}{Chen, Y.}, \bibinfo{author}{Somashekar, G.}, \bibinfo{author}{Ma, M.}, \bibinfo{author}{Simmhan, Y.}, \bibinfo{author}{Zhang, X.}, \bibinfo{author}{Mace, J.}, \bibinfo{author}{Vandevoorde, D.}, \bibinfo{author}{Las-Casas, P.}, \bibinfo{author}{Gupta, S.M.}, et~al., \bibinfo{year}{2024}.
\newblock \bibinfo{title}{Building ai agents for autonomous clouds: Challenges and design principles}, in: \bibinfo{booktitle}{Proceedings of the 2024 ACM Symposium on Cloud Computing}, pp. \bibinfo{pages}{99--110}.
\bibitem[{Tahir et~al.(2018)Tahir, Yamashita, Licorish, Dietrich and Counsell}]{tahir2018can}
\bibinfo{author}{Tahir, A.}, \bibinfo{author}{Yamashita, A.}, \bibinfo{author}{Licorish, S.}, \bibinfo{author}{Dietrich, J.}, \bibinfo{author}{Counsell, S.}, \bibinfo{year}{2018}.
\newblock \bibinfo{title}{Can you tell me if it smells? a study on how developers discuss code smells and anti-patterns in stack overflow}, in: \bibinfo{booktitle}{Proceedings of the 22nd International Conference on Evaluation and Assessment in Software Engineering 2018}, pp. \bibinfo{pages}{68--78}.
\bibitem[{Tian et~al.(2025)Tian, Luo, Du, Xian, Specht, Wang, Bi, Zhou, Kundu, Srinivasa et~al.}]{tian2025outlook}
\bibinfo{author}{Tian, F.}, \bibinfo{author}{Luo, A.}, \bibinfo{author}{Du, J.}, \bibinfo{author}{Xian, X.}, \bibinfo{author}{Specht, R.}, \bibinfo{author}{Wang, G.}, \bibinfo{author}{Bi, X.}, \bibinfo{author}{Zhou, J.}, \bibinfo{author}{Kundu, A.}, \bibinfo{author}{Srinivasa, J.}, et~al., \bibinfo{year}{2025}.
\newblock \bibinfo{title}{An outlook on the opportunities and challenges of multi-agent ai systems}.
\newblock \bibinfo{journal}{arXiv preprint arXiv:2505.18397} .
\bibitem[{Uddin et~al.(2021)Uddin, Sabir, Gu{\'e}h{\'e}neuc, Alam and Khomh}]{uddin2021empirical}
\bibinfo{author}{Uddin, G.}, \bibinfo{author}{Sabir, F.}, \bibinfo{author}{Gu{\'e}h{\'e}neuc, Y.G.}, \bibinfo{author}{Alam, O.}, \bibinfo{author}{Khomh, F.}, \bibinfo{year}{2021}.
\newblock \bibinfo{title}{An empirical study of iot topics in iot developer discussions on stack overflow}.
\newblock \bibinfo{journal}{Empirical Software Engineering} \bibinfo{volume}{26}, \bibinfo{pages}{121}.
\bibitem[{de~Witt(2025)}]{de2025open}
\bibinfo{author}{de~Witt, C.S.}, \bibinfo{year}{2025}.
\newblock \bibinfo{title}{Open challenges in multi-agent security: Towards secure systems of interacting ai agents}.
\newblock \bibinfo{journal}{arXiv preprint arXiv:2505.02077} .
\bibitem[{Woolson et~al.(1986)Woolson, Bean and Rojas}]{woolson1986sample}
\bibinfo{author}{Woolson, R.F.}, \bibinfo{author}{Bean, J.A.}, \bibinfo{author}{Rojas, P.B.}, \bibinfo{year}{1986}.
\newblock \bibinfo{title}{Sample size for case-control studies using cochran's statistic}.
\newblock \bibinfo{journal}{Biometrics} , \bibinfo{pages}{927--932}.
\bibitem[{Yang et~al.(2023)Yang, Zhang and Pan}]{yang2023understanding}
\bibinfo{author}{Yang, W.}, \bibinfo{author}{Zhang, C.}, \bibinfo{author}{Pan, M.}, \bibinfo{year}{2023}.
\newblock \bibinfo{title}{Understanding the topics and challenges of gpu programming by classifying and analyzing stack overflow posts}, in: \bibinfo{booktitle}{Proceedings of the 31st ACM Joint European Software Engineering Conference and Symposium on the Foundations of Software Engineering}, pp. \bibinfo{pages}{1444--1456}.
\bibitem[{Yang et~al.(2016)Yang, Lo, Xia, Wan and Sun}]{yang2016security}
\bibinfo{author}{Yang, X.L.}, \bibinfo{author}{Lo, D.}, \bibinfo{author}{Xia, X.}, \bibinfo{author}{Wan, Z.Y.}, \bibinfo{author}{Sun, J.L.}, \bibinfo{year}{2016}.
\newblock \bibinfo{title}{What security questions do developers ask? a large-scale study of stack overflow posts}.
\newblock \bibinfo{journal}{Journal of Computer Science and Technology} \bibinfo{volume}{31}, \bibinfo{pages}{910--924}.
\bibitem[{Zou et~al.(2025)Zou, Lin, Jones, Nowak, Dziemian, Winter, Grattan, Nathanael, Croft, Davies et~al.}]{zou2025security}
\bibinfo{author}{Zou, A.}, \bibinfo{author}{Lin, M.}, \bibinfo{author}{Jones, E.}, \bibinfo{author}{Nowak, M.}, \bibinfo{author}{Dziemian, M.}, \bibinfo{author}{Winter, N.}, \bibinfo{author}{Grattan, A.}, \bibinfo{author}{Nathanael, V.}, \bibinfo{author}{Croft, A.}, \bibinfo{author}{Davies, X.}, et~al., \bibinfo{year}{2025}.
\newblock \bibinfo{title}{Security challenges in ai agent deployment: Insights from a large scale public competition}.
\newblock \bibinfo{journal}{arXiv preprint arXiv:2507.20526} .

\end{thebibliography}
\end{document}